\documentclass[english]{iopart}
\usepackage[T1]{fontenc}
\usepackage[latin9]{inputenc}
\usepackage{amsbsy}
\usepackage{amssymb}
\usepackage{esint}

\makeatletter

\usepackage{iopams}
\usepackage{setstack}
\usepackage{multicol}
\usepackage{fullpage}
\usepackage{cite}
\usepackage{pgf,tikz,type1cm,fix-cm}

\makeatother

\usepackage{babel}
\begin{document}

\title{Time and dark matter from the\\
conformal symmetries of Euclidean space}

\author{Jeffrey S. Hazboun and James T. Wheeler}

\address{Utah State University, Physics Department, Logan, UT 84322}

\ead{\mailto{jeffrey.hazboun@gmail.com} \mailto{jim.wheeler@usu.edu}}
\begin{abstract}
The quotient of the conformal group of Euclidean 4-space by its Weyl
subgroup results in a geometry possessing many of the properties of
relativistic phase space, including both a natural symplectic form
and non-degenerate Killing metric. We show that the general solution
posesses orthogonal Lagrangian submanifolds, with the induced metric
and the spin connection on the submanifolds necessarily Lorentzian,
despite the Euclidean starting pont. By examining the structure equations
of the biconformal space in an orthonormal frame adapted to its phase
space properties, we also find that two new tensor fields exist in
this geometry, not present in Riemannian geometry. The first is a
combination of the Weyl vector with the scale factor on the metric,
and determines the timelike directions on the submanifolds. The second
comes from the components of the spin connection, symmetric with respect
to the new metric. Though this field comes from the spin connection
it transforms homogeneously. Finally, we show that in the absence
of conformal curvature or sources, the configuration space has geometric
terms equivalent to a perfect fluid and a cosmological constant. 
\end{abstract}

\submitto{\CQG}


\maketitle

\section{Introduction \label{sec:Historical-introduction}}

We develop a gauge theory based on the conformal group of a Euclidean
space, and show that its group properties necessarily lead to a Lorentzian
phase spacetime with \emph{vacuum} solutions carrying both a cosmological
constant and a cosmological perfect fluid as part of the generalized
Einstein tensor. In curved models, this geometric background may explain
or contribute to dark matter and dark energy. To emphasize the purely
geometric character of the construction, we give a description of
our use of the quotient manifold method for building gauge theories.
Our use of the conformal group, together with our choice of local
symmetry, lead to several structures not present in other related
gauge theories. Specifically, we show the generic presence of a symplectic
form, that there exists an induced metric from the non-degenerate
Killing form, demonstrate (but do not use) Kähler structure, and find
natural orthogonal, Lagrangian submanifolds. All of these properties
arise directly from group theory.

In the remainder of this introduction, we give a brief historical
overview of techniques leading up to, related to, or motivating our
own, then describe the layout of our presentation.

As mathematicians began studying the various incarnations of non-Euclidean
geometry, Klein started his Erlangen Program in 1872 as a way to classify
all forms of geometries that could be constructed using quotients
of groups. These \emph{homogeneous spaces} allowed for straightforward
classification of the spaces dependent on their symmetry properties.
Much of the machinery necessary to understand these spaces originated
with Cartan, beginning with his doctoral dissertation \cite{Cartan1910a}.
The classification of these geometries according to symmetry foreshadowed
gauge theory, the major tool that would be used by theoretical physicists
as the twentieth century continued. We will go into extensive detail
about how these methods are used in a modern context in section \ref{sec:Quotient-Manifold-Method}.
Most of the development, in modern language, can be found in \cite{Sharpe:1997a}. 

The use of symmetries to construct physical theories can be greatly
credited to Weyl's attempts at constructing a unified theory of gravity
and electromagnetism by adding dilatational symmetry to general relativity.
These attempts failed until Weyl looked at a $U(1)$ symmetry of the
action instead, thus constructing the first gauge theory of electromagnetism.
These efforts were extended to non-Abelian groups by Yang and Mills
\cite{Yang:1954ek}, including all $SU(n)$ and described by the Yang-Mills
action. The success of these theories as quantum pre-cursors inspired
relativists to try and construct general relativity as a gauge theory.
Utiyama \cite{Utiyama:1956p1235} looked at GR based on the the Lorentz
group, followed by Kibble \cite{Kibble:1961p1468} who first gauged
the Poincaré group to form general relativity. 

Standard approaches to gauge theory begin with a matter action globally
invariant under some symmetry group $\mathcal{H}$. This action generally
fails to be locally symmetric due to the derivatives of the fields,
but can be made locally invariant by introducing an $\mathcal{H}$-covariant
derivative. The connection fields used for this derivative are called
gauge fields. The final step is to make the gauge fields dynamical
by constructing their field strengths, which may be thought of as
curvatures of the connection, and including them in a modified action.

In the 1970's the success of the standard model and the growth of
supersymmetric gravity theories inspired physicists to extend the
symmetry used to construct a gravitational theory. MacDowell and Mansouri~\cite{MacDowell:1977p1566} obtained general relativity by gauging
the de Sitter or anti-de Sitter groups, and using a Wigner-Inönu contraction
to recover Poincaré symmetry. As a pre-cursor to supersymmetrizing
Weyl gravity, two groups \cite{CrispimRomao:1977hj,CrispimRomao:1978vf,Kaku:1977pa,Kaku:1977rk}
looked at a gravitational theory based on the conformal group, using
the Weyl curvature-squared action. These approaches are top-down,
in the sense that they often start with a physical matter action and
generalize to a local symmetry that leads to interactions. However,
as this work expanded, physicists started using the techniques of
Cartan and Klein to organize and develop the structures systematically.

In \cite{Neeman:1978p1517,Neeman:1978p1521} Ne'eman and Regge develop
what they refer to as the quotient manifold technique to construct
a gauge theory of gravity based on the Poincaré group. Theirs is the
first construction of a gravitational gauge theory that uses Klein
(homogeneous) spaces as generalized versions of tangent spaces, applying
methods developed by Cartan \cite{Kobayashi:1969p2015} to characterize
a more general geometry. In their 1982 papers \cite{Ivanov:1982p1172,Ivanov:1982p1201},
Ivanov and Niederle exhaustively considered quotients of the Poincaré,
de Sitter, anti-de Sitter and Lorentzian conformal groups ($ISO\left(3,1\right)$,
$SO\left(4,1\right)$, $SO\left(3,2\right)$ and $SO\left(4,2\right)$)
by various subgroups containing the Lorentz group.

There are a number of more recent implementations of Cartan geometry
in the modern literature. One good introduction is Wise's use of Cartan
methods to look at the MacDowell-Mansouri action \cite{Wise:2006sm}.
The ``waywiser'' approach of visualizing these geometries is advocated
strongly, and gives a clear geometric way of undertsanding Cartan
geometry. The use of Cartan techniques in \cite{Gryb:2012qt} to look
at the Chern-Simons action in $2+1$ dimensions provides a nice example
of the versatility of the method. This action can be viewed as having
either Minkowski, de Sitter or anti-de Sitter symmetry and Cartan
methods allow a straightforward characterization of the theory given
the various symmetries. The analysis is extended to look first at
the conformal representation of these groups on the Euclidean surfaces
of the theory ($2$-dimensional spatial slices). The authors then
look specifically at shape dynamics, which is found equivalent to
the case when the Chern-Simons action has de Sitter symmetry. Tractor
calculus is yet another example using a quotient of the conformal
group, in which the associated tensor bundles are based on a linear,
$(n+2)$-dim representation of the group. This is a distinct gauging
from the one we study here, but one studied in \cite{Wheeler:2013ora}. 

Our research focuses primarily on gaugings of the conformal group.
Initially motivated by a desire to understand the physical role of
local scale invariance, the growing prospects of twistor string formulations
of gravity \cite{ArkaniHamed:2012nw} elevate the importance of understanding
its low-energy limit, which is expected to be a conformal gauge theory
of gravity. Interestingly, there are two distinct ways to formulate
gravitational theories based on the conformal group, first identified
in \cite{Ivanov:1982p1172,Ivanov:1982p1201} and developed in \cite{Wheeler:2013ora,Wehner:1999p1653,Wheeler:1997pc}.
Both of these lead directly to scale-invariant general relativity.
This is surprising since the best known conformal gravity theory is
the fourth-order theory developed by Weyl \cite{Weyl1918a,Weyl1918b,Weyl1919,Weyl1921a,Weyl1923a}
and Bach \cite{Bach1920}. When a Palatini style variation is applied
to Weyl gravity, it becomes second-order, scale-invariant general
relativity \cite{Wheeler:2013ora}.

The second gauging of the conformal group identified in these works
is the \emph{biconformal} gauging. Leading to scale-invariant general
relativity formulated on a $2n$-dimensional symplectic manifold,
the approach took a novel twist for homogeneous spaces in \cite{Spencer:2008p167}.
There it is shown that, because the biconformal gauging leads to a
zero-signature manifold of doubled dimension, we can start with the
conformal symmetry of a non-Lorentzian space while still arriving
at spacetime gravity. We describe the resulting signature theorem
in detail below, and considerably strengthen its conclusions. In addition
to necessarily developing a direction of time from a Euclidean-signature
starting point, we show that these models give a group-theoretically
driven candidate for dark matter.

In the next Section, we describe the quotient manifold method in detail,
providing an example by applying it to the Poincaré group to produce
Cartan and Riemannian geometries. Then, in Section \ref{sec:Conformal-Quotients},
we apply the method to the conformal group in the two distinct ways
outlined above. The first, called the auxiliary gauging, reproduces
Weyl gravity. Focusing on the second, we identify a number of properties
posessed by the homogeneous space of the biconformal gauging. In Section
4, we digress to complete both gaugings by modifying the quotient
manifold and connections, then writing appropriate action functionals,
thereby establishing physical theories of gravity. We return to study
the homogeneous space of the biconformal gauging in Section 5, developing
the Maurer-Cartan structure equations in an adapted basis. Then, in
the next Section, we transform a known solution to the structure equations
into the adapted basis and identify the properties of the resulting
space. This reveals two previously unknown objects, one a tensor of
rank three, and the other a vector. In Section 6 we find the form
of the connection and basis forms when restricted to the configuration
and momentum submanifolds. This reveals the possibility of Riemannian
curvature of the submanifolds, even though the Cartan curvature of
the full space vanishes. Imposing the form of the solution, we find
the configuration space has a generalized Einstein tensor which contains
both a cosmological constant and cosmological dust in addition to
the usual Einstein tensor. Finally, we summarize our results.

\section{Quotient Manifold Method\label{sec:Quotient-Manifold-Method}}

We are interested in geometries \textendash{} ultimately spacetime
geometries \textendash{} which have continuous local symmetries. The
structure of such systems is that of a principal fiber bundle with
Lie group fibers. The quotient method starts with a Lie group, $\mathcal{G}$,
with the desired local symmetry as a proper Lie subgroup. To develop
the local properties any representation will give equivalent results,
so without loss of generality we assume a linear representation, i.e.
a vector space $\mathcal{V}^{n+2}$ on which $\mathcal{G}$ acts.
Typically this will be either a signature $\left(p,q\right)$ (pseudo-)Euclidean
space or the corresponding spinor space. This vector space is useful
for describing the larger symmetry group, but is only a starting point
and will not appear in the theory.

The quotient method, laid out below, is identical in many respects
to the approaches of \cite{Wise:2006sm,Gryb:2012qt}. The nice geometric
interpretation of using a Klein space in place of a tangent space
to both characterize a curved manifold and take advantage of its metric
structure are also among the motivations for using the quotient method.
In what follows not all the manifolds we look at will be interpreted
as spacetime, so we choose not to use the interpretation of a Klein
space moving around on spacetime in a larger ambient space. Rather
we directly generalize the homogeneous space to add curvatures. The
homogeneous space becomes a local model for a more general curved
space, similar to the way that $\mathbb{R}^{n}$ provides a local
model for an $n$-dim Riemannian manifold.

We include a concise introduction here, but the reader can find a
more detailed exposition in \cite{Sharpe:1997a}. Our intention is
to make it clear that our ultimate conclusions have rigorous roots
in group theory, rather than to present a comprehensive mathematical
description.

\subsection{Construction of a principal $\mathcal{H}$-bundle $\mathfrak{B}\left(\mathcal{G},\pi,\mathcal{H},M_{0}\right)$
with connection}

Consider a Lie group, $\mathcal{G}$, and a non-normal Lie subgroup,
$\mathcal{H}$, on which $\mathcal{G}$ acts effectively and transitively.
The quotient of these is a homogeneous manifold, $M_{0}$. The points
of $M_{0}$ are the left cosets, 
\[
g\mathcal{H}=\left\{ g'\mid g'=gh\: for\: some\: h\in\mathcal{H}\right\} 
\]
so there is a natural $1-1$ mapping $g\mathcal{H}\leftrightarrow\mathcal{H}$.
The cosets are disjoint from one another and together cover $\mathcal{G}$.
There is a projection, $\pi:\mathcal{G}\rightarrow M_{0}$, defined
by $\pi\left(g\right)=g\mathcal{H}\in M_{0}$. There is also a right
action of $\mathcal{G}$, $g\mathcal{H}\mathcal{G}$, given for all
elements of $\mathcal{G}$ by right multiplication.

Therefore, $\mathcal{G}$ is a principal $\mathcal{H}$-bundle, $\mathfrak{B}\left(\mathcal{G},\pi,\mathcal{H},M_{0}\right)$,
where the fibers are the left cosets. This is the mathematical object
required to carry a gauge theory of the symmetry group $\mathcal{H}$.
Let the dimension of $\mathcal{G}$ be $m$, the dimension of $\mathcal{H}$
be $k$. Then the dimension of the manifold is $n=m-k$ and we write
$M_{0}^{\left(n\right)}$. Choosing a gauge amounts to picking a cross-section
of this bundle, i.e., one point from each of these copies of $\mathcal{H}$.
Local symmetry amounts to dynamical laws which are independent of
the choice of cross-section.

Lie groups have a natural Cartan connection given by the one-forms,
$\boldsymbol{\xi}^{A}$, dual to the group generators, $G_{A}$. Rewriting
the Lie algebra in terms of these dual forms leads immediately to
the Maurer-Cartan structure equations,
\begin{equation}
\mathbf{d}\boldsymbol{\xi}^{A}=-\frac{1}{2}c_{\; BC}^{A}\boldsymbol{\xi}^{B}\wedge\boldsymbol{\xi}^{C}\label{Maurer-Cartan}
\end{equation}
where $c_{\; BC}^{A}$ are the group structure constants, and $\wedge$
is the wedge product. The integrability condition for this equation
follows from the Poincaré lemma, $\mathbf{d}^{2}=0$, and turns out
to be precisely the Jacobi identity. Therefore, the Maurer-Cartan
equations together with their integrability conditions are completely
equivalent to the Lie algebra of $\mathcal{G}$.

Let $\boldsymbol{\xi}^{a}$ (where $a=1,\ldots,k$) be the subset
of one-forms dual to the generators of the subgroup, $\mathcal{H}$.
Let the remaining independent forms be labeled $\boldsymbol{\chi}^{\alpha}$.
Then the $\boldsymbol{\xi}^{a}$ give a connection on the fibers while
the $\boldsymbol{\chi}^{\alpha}$ span the co-tangent spaces to $M_{0}^{\left(n\right)}$.
We denote the manifold with connection by $\mathcal{M}_{0}^{\left(n\right)}=\left(M_{0}^{\left(n\right)},\boldsymbol{\xi}^{A}\right)$.

\subsection{Cartan generalization}

For a gravity theory, we require in general a curved geometry, $\mathcal{M}^{\left(n\right)}$.
To achieve this, the quotient method allows us to generalize both
the connection and the manifold. Since the principal fiber bundle
from the quotient is a local direct product, this is not changed if
we allow a generalization of the manifold, $M_{0}^{\left(n\right)}\rightarrow M^{\left(n\right)}$.
We will not consider such topological issues here. Generalizing the
connection is more subtle. If we change $\boldsymbol{\xi}^{A}=\left(\boldsymbol{\xi}^{a},\boldsymbol{\chi}^{\alpha}\right)$
to a new connection $\boldsymbol{\xi}^{A}\rightarrow\boldsymbol{\omega}^{A},\boldsymbol{\xi}^{a}\rightarrow\boldsymbol{\omega}^{a},\boldsymbol{\chi}^{\alpha}\rightarrow\boldsymbol{\omega}^{\alpha}$
arbitrarily, the Maurer-Cartan equation is altered to
\[
\mathbf{d}\boldsymbol{\omega}^{A}=-\frac{1}{2}c_{\; BC}^{A}\boldsymbol{\omega}^{B}\wedge\boldsymbol{\omega}^{C}+\boldsymbol{\Omega}^{A}
\]
where $\boldsymbol{\Omega}^{A}$ is a $2$-form determined by the
choice of the new connection. We need restrictions on $\boldsymbol{\Omega}^{A}$
so that it represents curvature of the geometry $\mathcal{M}^{\left(n\right)}=\left(M^{\left(n\right)},\boldsymbol{\omega}^{A}\right)$
and not of the full bundle $\mathfrak{B}$. We restrict $\boldsymbol{\Omega}^{A}$
by requiring it to be independent of lifting, i.e., horizontality
of the curvature.

To define horizontality, recall that the integral of the connection
around a closed curve in the bundle is given by the integral of $\boldsymbol{\Omega}^{A}$
over any surface bounded by the curve. We require this integral to
be independent of lifting, i.e., horizontal. It is easy to show that
this means that the two-form basis for the curvatures $\boldsymbol{\Omega}^{A}$
cannot include any of the one-forms, $\boldsymbol{\omega}^{a}$, that
span the fiber group, $\mathcal{H}$. With the horizontality condition,
the curvatures take the simpler form
\[
\boldsymbol{\Omega}^{A}=\frac{1}{2}\Omega_{\;\alpha\beta}^{A}\boldsymbol{\omega}^{\alpha}\wedge\boldsymbol{\omega}^{\beta}
\]
More general curvatures than this will destroy the homogeneity of
the fibers, so we would no longer have a \emph{principal} $\mathcal{H}$-bundle.

In addition to horizontality, we require integrability. Again using
the Poincaré lemma, $\mathbf{d}^{2}\boldsymbol{\omega}^{A}\equiv0$,
we always find a term $\frac{1}{2}c_{\; B[C}^{A}c_{\; DE]}^{B}\boldsymbol{\omega}^{C}\wedge\boldsymbol{\omega}^{D}\wedge\boldsymbol{\omega}^{E}$
which vanishes by the Jacobi identity, $c_{\; B[C}^{A}c_{\; DE]}^{B}\equiv0$,
while the remaining terms give the general form of the Bianchi identities,
\[
\mathbf{d}\boldsymbol{\Omega}^{A}+c_{\; BC}^{A}\boldsymbol{\omega}^{B}\wedge\boldsymbol{\Omega}^{C}=0
\]

\subsection{Example: Pseudo-Riemannian manifolds}

To see how this works in a familiar example, consider the construction
of the pseudo-Riemannian spacetimes used in general relativity, for
which we take the quotient of the $10$-dim Poincaré group by its
$6$-dim Lorentz subgroup. The result is a principal Lorentz bundle
over $\mathbb{R}^{4}$. Writing the one-forms dual to the Lorentz
$\left(M_{\; b}^{a}\right)$ and translation $\left(P_{a}\right)$
generators as $\boldsymbol{\xi}_{\; b}^{a}$ and $\boldsymbol{\omega}^{a}$,
respectively, the ten Maurer-Cartan equations are
\begin{eqnarray*}
\mathbf{d}\boldsymbol{\xi}_{\; b}^{a} & = & \boldsymbol{\xi}_{\; b}^{c}\wedge\boldsymbol{\xi}_{\; c}^{a}\\
\mathbf{d}\boldsymbol{\omega}^{a} & = & \boldsymbol{\omega}^{b}\wedge\boldsymbol{\xi}_{\; b}^{a}
\end{eqnarray*}
Notice that the first describes a pure gauge spin connection, $\mathbf{d}\boldsymbol{\xi}_{\; b}^{a}=-\bar{\Lambda}_{\; b}^{c}\mathbf{d}\Lambda_{\; c}^{a}$
where $\Lambda_{\; c}^{a}$ is a local Lorentz transformation. Therefore,
there exists a local Lorentz gauge such that $\boldsymbol{\xi}_{\; b}^{a}=0$.
The second equation then shows the existence of an exact orthonormal
frame, which tells us that the space is Minkowski.

Now generalize the geometry, $\left(M_{0}^{4},\boldsymbol{\xi}^{A}\right)\rightarrow\left(M^{4},\boldsymbol{\omega}^{A}\right)$
where $M_{0}^{4}=\mathbb{R}^{4}$ and we denote the new connection
forms by $\boldsymbol{\omega}^{A}=\left(\boldsymbol{\omega}_{\; b}^{a},\mathbf{e}^{b}\right)$.
In the structure equations, this leads to the presence of ten curvature
$2$-forms,
\begin{eqnarray*}
\mathbf{d}\boldsymbol{\omega}_{\; b}^{a} & = & \boldsymbol{\omega}_{\; b}^{c}\wedge\boldsymbol{\omega}_{\; c}^{a}+\mathbf{R}_{\; b}^{a}\\
\mathbf{d}\mathbf{e}^{a} & = & \mathbf{e}^{b}\wedge\boldsymbol{\omega}_{\; b}^{a}+\mathbf{T}^{a}
\end{eqnarray*}
Since the $\boldsymbol{\omega}_{\; b}^{a}$ span the Lorentz subgroup,
horizontality is accomplished by restricting the curvatures to
\begin{eqnarray*}
\mathbf{R}_{\; b}^{a} & = & \frac{1}{2}R_{\; bcd}^{a}\mathbf{e}^{c}\wedge\mathbf{e}^{d}\\
\mathbf{T}^{a} & = & \frac{1}{2}T_{\; bc}^{a}\mathbf{e}^{b}\wedge\mathbf{e}^{c}
\end{eqnarray*}
that is, there are no terms such as, for example, $\frac{1}{2}R_{\; b\; de}^{a\; c}\boldsymbol{\omega}_{\; c}^{d}\wedge\mathbf{e}^{e}$
or $\frac{1}{2}T_{\; b\; d}^{a\; c\; e}\boldsymbol{\omega}_{\; c}^{b}\wedge\boldsymbol{\omega}_{\; e}^{d}$.
Finally, integrability is guaranteed by the pair of Bianchi identities,
\begin{eqnarray*}
\mathbf{d}\mathbf{R}_{\; b}^{a}+\mathbf{R}_{\; b}^{c}\wedge\boldsymbol{\omega}_{\; c}^{a}-\mathbf{R}_{\; c}^{a}\wedge\boldsymbol{\omega}_{\; b}^{c} & = & 0\\
\mathbf{d}\mathbf{T}^{a}+\mathbf{T}^{b}\wedge\boldsymbol{\omega}_{\; b}^{a}+\mathbf{e}^{b}\wedge\mathbf{R}_{\; b}^{a} & = & 0
\end{eqnarray*}
By looking at the transformation of $\mathbf{R}_{\; b}^{a}$ and $\mathbf{T}^{a}$
under local Lorentz transformations, we find that despite originating
as components of a single Poincaré-valued curvature, they are independent
Lorentz tensors. The translations of the Poincaré symmetry were broken
when we curved the base manifold (see \cite{Kibble:1961p1468,Neeman:1978p1517,Neeman:1978p1521},
but note that Kibble effectively uses a $14$-dimensional bundle,
whereas ours and related approaches require only $10$-dim). We recognize
$\mathbf{R}_{\; b}^{a}$ and $\mathbf{T}^{a}$ as the Riemann curvature
and the torsion two-forms, respectively. Since the torsion is an independent
tensor under the fiber group, it is consistent to consider the subclass
of Riemannian geometries, $\mathbf{T}^{a}=0$. Alternatively (see
Sec. \ref{sec: gravity} below), vanishing torsion follows from the
Einstein-Hilbert action.

With vanishing torsion, the quotient method has resulted in the usual
solder form, $\mathbf{e}^{a}$, and related metric-compatible spin
connection, $\boldsymbol{\omega}_{\; b}^{a}$,
\[
\mathbf{d}\mathbf{e}^{a}-\mathbf{e}^{b}\wedge\boldsymbol{\omega}_{\; b}^{a}=0,
\]
the expression for the Riemannian curvature in terms of these,
\[
\mathbf{R}_{\; b}^{a}=\mathbf{d}\boldsymbol{\omega}_{\; b}^{a}-\boldsymbol{\omega}_{\; b}^{c}\wedge\boldsymbol{\omega}_{\; c}^{a},
\]
and the first and second Bianchi identities,
\begin{eqnarray*}
\mathbf{e}^{b}\wedge\mathbf{R}_{\; b}^{a} & = & 0\\
\mathbf{D}\mathbf{R}_{\; b}^{a} & = & 0.
\end{eqnarray*}
This is a complete description of the class of Riemannian geometries.

Many further examples were explored by Ivanov and Niederle \cite{Ivanov:1982p1172,Ivanov:1982p1201}.

\section{Quotients of the conformal group \label{sec:Conformal-Quotients}}

\subsection{General properties of the conformal group}

Physically, we are interested in measurements of relative magnitudes,
so the relevant group is the conformal group, $\mathcal{C}$, of compactified
$\mathbb{R}^{n}$. The one-point compactification at infinity allows
a global definition of inversion, with translations of the point at
infinity defining the special conformal transformation. Then $\mathcal{C}$
has a real linear representation in $n+2$ dimensions, $\mathcal{V}^{n+2}$
(alternatively we could choose the complex representation $\mathbb{C}^{2^{[\left(n+2\right)/2]}}$
for $Spin\left(p+1,q+1\right)$). The isotropy subgroup of $\mathbb{R}^{n}$
is the rotations, $SO\left(p,q\right)$, together with dilatations.
We call this subgroup the homogeneous Weyl group, $\mathcal{W}$ and
require our fibers to contain it. There are then only three allowed
subgroups: $\mathcal{W}$ itself; the inhomogeneous Weyl group, $\mathcal{I}\mathcal{W}$,
found by appending the translations; and $\mathcal{W}$ together with
special conformal transformations, isomorphic to $\mathcal{I}\mathcal{W}$.
The quotient of the conformal group by either inhomogeneous Weyl group,
called the \emph{auxiliary gauging}, leads most naturally to Weyl
gravity {[}for a review, see \cite{Wheeler:2013ora}{]}. We concern
ourselves with the only other meaningful conformal quotient, the \emph{biconformal
gauging}: the principal $\mathcal{W}$-bundle formed by the quotient
of the conformal group by its Weyl subgroup. To help clarify the method
and our model, it is useful to consider both these gaugings.

All parts of this construction work for any $\left(p,q\right)$ with
$n=p+q$. The conformal group is then $SO\left(p+1,q+1\right)$ (or
$Spin\left(p+1,q+1\right)$ for the twistor representation). The Maurer-Cartan
structure equations are immediate. In addition to the $\frac{n\left(n-1\right)}{2}$
generators $M_{\;\beta}^{\alpha}$ of $SO\left(p,q\right)$ and $n$
translational generators $P_{\alpha}$, there are $n$ generators
of translations of a point at infinity (``special conformal transformations'')
$K^{\alpha}$, and a single dilatational generator $D$. Dual to these,
we have the connections $\xi_{\;\beta}^{\alpha},\boldsymbol{\chi}^{\alpha},\boldsymbol{\pi}_{\alpha},$$\boldsymbol{\delta}$,
respectively. Substituting the structure constants into the Maurer-Cartan
dual form of the Lie algebra, eq.(\ref{Maurer-Cartan}) gives
\begin{eqnarray}
\mathbf{d}\boldsymbol{\xi}_{\;\beta}^{\alpha} & = & \boldsymbol{\xi}_{\;\beta}^{\mu}\wedge\boldsymbol{\xi}_{\;\mu}^{\alpha}+2\Delta_{\nu\beta}^{\alpha\mu}\boldsymbol{\pi}_{\mu}\wedge\boldsymbol{\chi}^{\nu}\label{Homogeneous Spin Connection}\\
\mathbf{d}\boldsymbol{\chi}^{\alpha} & = & \boldsymbol{\chi}^{\beta}\wedge\boldsymbol{\xi}_{\;\beta}^{\alpha}+\boldsymbol{\delta}\wedge\boldsymbol{\chi}^{\alpha}\label{Homogeneous Solder Form}\\
\mathbf{d}\boldsymbol{\pi}_{\alpha} & = & \boldsymbol{\xi}_{\;\alpha}^{\beta}\wedge\boldsymbol{\pi}_{\beta}-\boldsymbol{\delta}\wedge\boldsymbol{\pi}_{\alpha}\label{Homogeneous Co-Solder Form}\\
\mathbf{d}\boldsymbol{\delta} & = & \boldsymbol{\chi}^{\alpha}\wedge\boldsymbol{\pi}_{\alpha}\label{Homogeneous Weyl form}
\end{eqnarray}
where $\Delta_{\nu\beta}^{\alpha\mu}\equiv\frac{1}{2}\left(\delta_{\nu}^{\alpha}\delta_{\beta}^{\mu}-\delta^{\alpha\mu}\delta_{\nu\beta}\right)$
antisymmetrizes \emph{with respect to the original $\left(p,q\right)$
metric}, $\delta_{\mu\nu}=diag\left(1,\ldots,1,-1,\ldots,-1\right)$.
These equations, which are the same regardless of the gauging chosen,
describe the Cartan connection on the conformal group manifold. Before
proceeding to the quotients, we note that the conformal group has
a nondegenerate Killing form, 
\[
K_{AB}\equiv tr\left(G_{A}G_{B}\right)=c_{\; AD}^{C}c_{\; BC}^{D}=\left(\begin{array}{cccc}
\Delta_{db}^{ac}\\
 & 0 & \delta_{b}^{a}\\
 & \delta_{b}^{a} & 0\\
 &  &  & 1
\end{array}\right)
\]
This provides a metric on the conformal Lie algebra. As we show below,
when restricted to $\mathcal{M}_{0}$, it may or may not remain nondegenerate,
depending on the quotient.

Finally, we note that the conformal group is invariant under inversion.
Within the Lie algebra, this manifests itself as the interchange between
the translations and special conformal transformations $P_{\alpha}\leftrightarrow\delta_{\alpha\beta}K^{\beta}$
along with the interchange of conformal weights, $D\rightarrow-D$.
The corresponding transformation of the connection forms, is easily
seen to leave eqs.(\ref{Homogeneous Spin Connection})-(\ref{Homogeneous Weyl form})
invariant. In the biconformal gauging, below, we show that this symmetry
leads to a Kähler structure.

\subsection{Curved generalizations}

In this sub-Section and Section \ref{sec: gravity} we will complete
the development of the curved auxiliary and biconformal geometries
and show how one can easily construct actions with the curvatures.
In this sub-Section, we construct the two possible fiber bundles,
$\mathcal{C}/\mathcal{H}$ where $\mathcal{W}\subseteq\mathcal{H}$.
For each, we carry out the generalization of the manifold and connection.
The results in this sub-Section depend only on whether the local symmetry
is $\mathcal{H}=\mathcal{I}\mathcal{W}$ or $\mathcal{H}=\mathcal{W}$.
In Section \ref{sec:PropertiesHomogeneousBCS} and Section \ref{sec:RiemannianStrucInBCS}
we will return to the un-curved case to present a number of new calculations
characterizing the homogenous space formed from the biconformal gauging. 

The first sub-Section below describes the \emph{auxiliary gauging},
given by the quotient of the conformal group by the inhomogeneous
Weyl group, $\mathcal{I}\mathcal{W}$.

Since $\mathcal{I}\mathcal{W}$ is a parabolic subgroup of the conformal
group, the resulting quotient can be considered a \emph{tractor space},
for which there are numerous results \cite{Cap:2009a}. \emph{Tractor
calculus} is a version of the auxiliary gauging where the original
conformal group is tensored with $\mathbb{R}^{\left(p+1,q+1\right)}$.
This allows for a linear representation of the conformal group with
$\left(n+2\right)$-dimensional tensorial (physical) entities called
\emph{tractors}. This linear representation, first introduced by Dirac
\cite{Dirac1936a}, makes a number of calculations much easier and
also allows for straightforward building of tensors of any rank. The
main physical differences stem from the use of Dirac's action, usually
encoded as the scale tractor squared in the $n+2$-dimensional linear
representation, instead of the Weyl action we introduce in Sec \ref{sec: gravity}. 

In sub-Section \ref{sub:The-biconformal-gauging} below, we quotient
by the homogeneous Weyl group, giving the \emph{biconformal gauging}.
This is \emph{not} a parabolic quotient and therefore represents a
less conventional option which turns out to have a number of rich
structures not present in the auxiliary gauging. The biconformal gauging
will occupy our attention for the bulk of our subsequent discussion.

\subsubsection{The auxiliary gauging: $\mathcal{H}=\mathcal{I}\mathcal{W}$}

Given the quotient $\mathcal{C}/\mathcal{I}\mathcal{W}$, the one-forms
$\left(\boldsymbol{\xi}_{\;\beta}^{\alpha},\boldsymbol{\pi}_{\mu},\boldsymbol{\delta}\right)$
span the $\mathcal{I}\mathcal{W}$-fibers, with $\boldsymbol{\beta}^{\alpha}$
spanning the co-tangent space of the remaining $n$ independent directions.
This means that $\mathcal{M}_{0}^{\left(n\right)}$ has the same dimension,
$n$, as the original space. Generalizing the connection, we replace
$\left(\boldsymbol{\xi}_{\;\beta}^{\alpha},\boldsymbol{\chi}^{\alpha},\boldsymbol{\pi}_{\alpha},\boldsymbol{\delta}\right)\rightarrow\left(\boldsymbol{\omega}_{\;\beta}^{\alpha},\mathbf{e}^{\alpha},\boldsymbol{\omega}_{\alpha},\boldsymbol{\omega}\right)$
and the Cartan equations now give the Cartan curvatures in terms of
the new connection forms,
\begin{eqnarray}
\mathbf{d}\boldsymbol{\omega}_{\;\beta}^{\alpha} & = & \boldsymbol{\omega}_{\;\beta}^{\mu}\wedge\boldsymbol{\omega}_{\;\mu}^{\alpha}+2\Delta_{\nu\beta}^{\alpha\mu}\boldsymbol{\omega}_{\mu}\wedge\boldsymbol{\omega}^{\nu}+\boldsymbol{\Omega}_{\;\beta}^{\alpha}\label{Aux spin connection}\\
\mathbf{d}\mathbf{e}^{\alpha} & = & \mathbf{e}^{\beta}\wedge\boldsymbol{\omega}_{\;\beta}^{\alpha}+\boldsymbol{\omega}\wedge\mathbf{e}^{\alpha}+\mathbf{T}^{\alpha}\label{Aux solder form}\\
\mathbf{d}\boldsymbol{\omega}_{\alpha} & = & \boldsymbol{\omega}_{\;\alpha}^{\beta}\wedge\boldsymbol{\omega}_{\beta}-\boldsymbol{\omega}\wedge\boldsymbol{\omega}_{\alpha}+\mathbf{S}_{\alpha}\label{Aux special conformal}\\
\mathbf{d}\boldsymbol{\omega} & = & \boldsymbol{\omega}^{\alpha}\wedge\boldsymbol{\omega}_{\alpha}+\boldsymbol{\Omega}\label{Aux dilatation}
\end{eqnarray}
Up to local gauge transformations, the curvatures depend only on the
$n$ non-vertical forms, $\mathbf{e}^{\alpha}$, so the curvatures
are similar to what we find in an $n$-dim Riemannian geometry. For
example, the $SO\left(p,q\right)$ piece of the curvature takes the
form $\boldsymbol{\Omega}_{\;\beta}^{\alpha}=\frac{1}{2}\Omega_{\;\beta\mu\nu}^{\alpha}\mathbf{e}^{\alpha}\wedge\mathbf{e}^{\beta}$.
The coefficients have the same number of degrees of freedom as the
Riemannian curvature of an $n$-dim Weyl geometry.

Finally, each of the curvatures has a corresponding Bianchi identity,
to guarantee integrability of the modified structure equations,
\begin{eqnarray}
0 & = & \mathbf{D}\boldsymbol{\Omega}_{\;\beta}^{\alpha}+2\Delta_{\nu\beta}^{\alpha\mu}\left(\boldsymbol{\Omega}_{\mu}\wedge\boldsymbol{\omega}^{\nu}-\boldsymbol{\omega}_{\mu}\wedge\boldsymbol{\Omega}^{\nu}\right)\label{Bianchi for rotations}\\
0 & = & \mathbf{D}\mathbf{T}^{\alpha}-\mathbf{e}^{\beta}\wedge\boldsymbol{\Omega}_{\;\beta}^{\alpha}+\boldsymbol{\Omega}\wedge\mathbf{e}^{\alpha}\label{Bianchi for torsion}\\
0 & = & \mathbf{D}\mathbf{S}_{\alpha}+\boldsymbol{\Omega}_{\;\beta}^{\alpha}\wedge\boldsymbol{\omega}_{\beta}-\boldsymbol{\omega}_{\alpha}\wedge\boldsymbol{\Omega}\label{Bianchi for co-torsion}\\
0 & = & \mathbf{D}\boldsymbol{\Omega}+\mathbf{T}^{\alpha}\wedge\boldsymbol{\omega}_{\alpha}-\boldsymbol{\omega}^{\alpha}\wedge\mathbf{S}_{\alpha}\label{Bianchi for dilatations}
\end{eqnarray}
where $D$ is the Weyl covariant derivative,
\begin{eqnarray*}
\mathbf{D}\boldsymbol{\Omega}_{\;\beta}^{\alpha} & = & \mathbf{d}\boldsymbol{\Omega}_{\;\beta}^{\alpha}+\boldsymbol{\Omega}_{\;\beta}^{\mu}\wedge\boldsymbol{\omega}_{\;\mu}^{\alpha}-\boldsymbol{\Omega}_{\;\mu}^{\alpha}\wedge\boldsymbol{\omega}_{\;\beta}^{\mu}\\
\mathbf{D}\mathbf{T}^{\alpha} & = & \mathbf{d}\mathbf{T}^{\alpha}+\mathbf{T}^{\beta}\wedge\boldsymbol{\omega}_{\;\beta}^{\alpha}-\boldsymbol{\omega}\wedge\mathbf{T}^{\alpha}\\
\mathbf{D}\mathbf{S}_{\alpha} & = & \mathbf{d}\mathbf{S}_{\alpha}-\boldsymbol{\omega}_{\;\alpha}^{\beta}\wedge\mathbf{S}_{\beta}+\mathbf{S}_{\alpha}\wedge\boldsymbol{\omega}\\
\mathbf{D}\boldsymbol{\Omega} & = & \mathbf{d}\boldsymbol{\Omega}
\end{eqnarray*}
Equations eq.(\ref{Aux spin connection}-\ref{Aux dilatation}) give
the curvature two-forms in terms of the connection forms. We have
therefore constructed an $n$-dim geometry based on the conformal
group with local $\mathcal{IW}$ symmetry.

We note no additional special properties of these geometries from
the group structure. In particular, the restriction (in square brackets,
$\left[\;\right]$, below) of the Killing metric, $K_{AB}$, to $\mathcal{M}^{\left(n\right)}$
vanishes identically,
\[
\left.\left(\begin{array}{cccc}
\Delta_{db}^{ac}\\
 & \left[0\right] & \delta_{b}^{a}\\
 & \delta_{b}^{a} & 0\\
 &  &  & 1
\end{array}\right)\right|_{\mathcal{M}^{\left(n\right)}}=\left(\begin{array}{c}
0\end{array}\right)_{n\times n},
\]
so there is no induced metric on the spacetime manifold. We may add
the usual metric by hand, of course, but our goal here is to find
those properties which are intrinsic to the underlying group structures.

\subsubsection{The biconformal gauging: $\mathcal{H}=\mathcal{W}$ \label{sub:The-biconformal-gauging}}

We next consider the biconformal gauging, first considered by Ivanov
and Niederle \cite{Ivanov:1982p1201}, given by the quotient of the
conformal group by its Weyl subgroup. The resulting geometry has been
shown to contain the structures of general relativity \cite{Wheeler:1997pc,Wehner:1999p1653}.

Given the quotient $\mathcal{C}/\mathcal{W}$, the one-forms $\left(\boldsymbol{\xi}_{\;\beta}^{\alpha},\boldsymbol{\delta}\right)$
span the $\mathcal{W}$-fibers, with $\left(\boldsymbol{\chi}^{\alpha},\boldsymbol{\pi}_{\alpha}\right)$
spanning the remaining $2n$ independent directions. This means that
$\mathcal{M}_{0}^{\left(2n\right)}$ has twice the dimension of the
original compactified $\mathbb{R}^{\left(n\right)}$. Generalizing,
we replace $\left(\boldsymbol{\xi}_{\;\beta}^{\alpha},\boldsymbol{\chi}^{\alpha},\boldsymbol{\pi}_{\alpha},\boldsymbol{\delta}\right)\rightarrow\left(\boldsymbol{\omega}_{\;\beta}^{\alpha},\boldsymbol{\omega}^{\alpha},\boldsymbol{\omega}_{\alpha},\boldsymbol{D}\right)$
and the modified structure equations \emph{appear} identical to eqs.(\ref{Aux spin connection}-\ref{Aux dilatation}).
However, the curvatures now depend on the $2n$ non-vertical forms,
$\left(\boldsymbol{\omega}^{\alpha},\boldsymbol{\omega}_{\alpha}\right)$,
so there are far more components than for an $n$-dim Riemannian geometry.
For example,
\[
\boldsymbol{\Omega}_{\;\beta}^{\alpha}=\frac{1}{2}\Omega_{\;\beta\mu\nu}^{\alpha}\boldsymbol{\omega}^{\mu}\wedge\boldsymbol{\omega}^{\nu}+\Omega_{\;\beta\;\nu}^{\alpha\;\mu}\boldsymbol{\omega}_{\mu}\wedge\boldsymbol{\omega}^{\nu}+\frac{1}{2}\Omega_{\;\beta}^{\alpha\;\mu\nu}\boldsymbol{\omega}_{\mu}\wedge\boldsymbol{\omega}_{\nu}
\]
The coefficients of the pure terms, $\Omega_{\;\beta\mu\nu}^{\alpha}$
and $\Omega_{\;\beta}^{\alpha\;\mu\nu}$ each have the same number
of degrees of freedom as the Riemannian curvature of an $n$-dim Weyl
geometry, while the cross-term coefficients $\Omega_{\;\beta\;\nu}^{\alpha\;\mu}$
have more, being asymmetric on the final two indices.

For our purpose, it is important to notice that the spin connection,
$\boldsymbol{\xi}_{\;\beta}^{\alpha}$, is antisymmetric with respect
to the original $\left(p,q\right)$ metric, $\delta_{\alpha\beta}$,
in the sense that 
\[
\boldsymbol{\xi}_{\;\beta}^{\alpha}=-\delta^{\alpha\mu}\delta_{\beta\nu}\boldsymbol{\xi}_{\;\mu}^{\nu}
\]
It is crucial to note that $\boldsymbol{\omega}_{\;\beta}^{\alpha}$
retains this property, $\boldsymbol{\omega}_{\;\beta}^{\alpha}=-\delta^{\alpha\mu}\delta_{\beta\nu}\boldsymbol{\omega}_{\;\mu}^{\nu}$.
This expresses metric compatibility with the $SO\left(p,q\right)$-covariant
derivative, since it implies
\[
\mathbf{D}\delta_{\alpha\beta}\equiv\mathbf{d}\delta_{\alpha\beta}-\delta_{\mu\beta}\boldsymbol{\omega}_{\;\alpha}^{\mu}-\delta_{\alpha\mu}\boldsymbol{\omega}_{\;\beta}^{\mu}=0
\]
Therefore, the curved generalization has a connection which is compatible
with a locally $\left(p,q\right)$-metric. This relationship is general.
If $\kappa_{\alpha\beta}$ is any metric, its compatible spin connection
will satisfy $\boldsymbol{\omega}_{\;\beta}^{\alpha}=-\kappa^{\alpha\mu}\kappa_{\beta\nu}\boldsymbol{\omega}_{\;\mu}^{\nu}$.
Since we also have local scale symmetry, the full covariant derivative
we use will also include a Weyl vector term.

The Bianchi identities, written as $2$-forms, also \emph{appear}
the same as eqs.(\ref{Bianchi for rotations}-\ref{Bianchi for dilatations}),
but expand into more components.

In the conformal group, translations and special conformal transformations
are related by inversion. Indeed, a special conformal tranformation
is a translation centered at the point at infinity instead of the
origin. Because the biconformal gauging maintains the symmetry between
translations and special conformal transformations, it is useful to
name the corresponding connection forms and curvatures to reflect
this. Therefore, the biconformal basis will be described as the solder
form and the co-solder form, and the corresponding curvatures as the
torsion and co-torsion. Thus, when we speak of ``torsion-free biconformal
space'' we do not imply that the co-torsion (Cartan curvature of
the co-solder form) vanishes. In phase space interpretations, the
solder form is taken to span the cotangent spaces of the spacetime
manifold, while the co-solder form is taken to span the cotangent
spaces of the momentum space. The opposite convention is equally valid.

Unlike other quotient manifolds arising in conformal gaugings, the
biconformal quotient manifold posesses natural invariant structures.
The first is the restriction of the Killing metric, which is now non-degenerate,
\[
\left.\left(\begin{array}{ccc}
\Delta_{db}^{ac}\\
 & \left[\begin{array}{cc}
0 & \delta_{b}^{a}\\
\delta_{b}^{a} & 0
\end{array}\right]\\
 &  & 1
\end{array}\right)\right|_{\mathcal{M}^{\left(2n\right)}}=\left(\begin{array}{cc}
0 & \delta_{b}^{a}\\
\delta_{b}^{a} & 0
\end{array}\right)_{2n\times2n},
\]
and this gives an inner product for the basis,

\begin{equation}
\left[\begin{array}{cc}
\left\langle \boldsymbol{\omega}^{\alpha},\boldsymbol{\omega}^{\beta}\right\rangle  & \left\langle \boldsymbol{\omega}^{\alpha},\boldsymbol{\omega}_{\beta}\right\rangle \\
\left\langle \boldsymbol{\omega}_{\alpha},\boldsymbol{\omega}^{\beta}\right\rangle  & \left\langle \boldsymbol{\omega}_{\alpha},\boldsymbol{\omega}_{\beta}\right\rangle 
\end{array}\right]\equiv\left[\begin{array}{cc}
0 & \delta_{\beta}^{\alpha}\\
\delta_{\alpha}^{\beta} & 0
\end{array}\right]\label{eq:KillingMetric-1}
\end{equation}
This metric remains unchanged by the generalization to curved base
manifolds.

The second natural invariant property is the generic presence of a
symplectic form. The original fiber bundle always has this, because
the structure equation, eq.(\ref{Homogeneous Weyl form}), shows that
$\boldsymbol{\chi}^{\alpha}\wedge\boldsymbol{\pi}_{\alpha}$ is exact
hence closed, $\mathbf{d}^{2}\boldsymbol{\omega}=0$, while it is
clear that the two-form product is non-degenerate because $\left(\boldsymbol{\chi}^{\alpha},\boldsymbol{\pi}_{\alpha}\right)$
together span $\mathcal{M}_{0}^{\left(2n\right)}$. Moreover, the
symplectic form is canonical,
\begin{eqnarray*}
\left[\Omega\right]_{AB} & = & \left[\begin{array}{cc}
0 & \delta_{\alpha}^{\beta}\\
-\delta_{\beta}^{\alpha} & 0
\end{array}\right]
\end{eqnarray*}
so that $\boldsymbol{\chi}^{\alpha}$ and $\boldsymbol{\pi}_{\alpha}$
are canonically conjugate. The symplectic form persists for the $2$-form,
$\boldsymbol{\omega}^{\alpha}\wedge\boldsymbol{\omega}_{\alpha}+\boldsymbol{\Omega}$,
as long as it is non-degenerate, so curved biconformal spaces are
generically symplectic.

Next, we consider the effect of inversion symmetry. As a $\left(\begin{array}{c}
1\\
1
\end{array}\right)$ tensor, the basis interchange takes the form
\[
I_{\; B}^{A}\boldsymbol{\chi}^{B}=\left(\begin{array}{cc}
0 & \delta^{\alpha\nu}\\
\delta_{\beta\mu} & 0
\end{array}\right)\left(\begin{array}{c}
\boldsymbol{\chi}^{\mu}\\
\boldsymbol{\pi}_{\nu}
\end{array}\right)=\left(\begin{array}{c}
\delta^{\alpha\nu}\boldsymbol{\pi}_{\nu}\\
\delta_{\beta\mu}\boldsymbol{\chi}^{\mu}
\end{array}\right)
\]
In order to interchange conformal weights, $I_{\; B}^{A}$ must anticommute
with the conformal weight operator, which is given by
\[
W_{\; B}^{A}\boldsymbol{\chi}^{B}=\left(\begin{array}{cc}
\delta_{\mu}^{\alpha} & 0\\
0 & -\delta_{\beta}^{\nu}
\end{array}\right)\left(\begin{array}{c}
\boldsymbol{\chi}^{\mu}\\
\boldsymbol{\pi}_{\nu}
\end{array}\right)=\left(\begin{array}{c}
+\boldsymbol{\chi}^{\alpha}\\
-\boldsymbol{\pi}_{\beta}
\end{array}\right)
\]
This is the case: we easily check that $\left\{ I,W\right\} _{\; B}^{A}=I_{\; C}^{A}W_{\; B}^{C}+W_{\; C}^{A}I_{\; B}^{C}=0$.
The commutator gives a new object,
\begin{eqnarray*}
J_{\; B}^{A} & \equiv & \left[I,W\right]_{\; B}^{A}=\left(\begin{array}{cc}
0 & -\delta^{\alpha\beta}\\
\delta_{\alpha\beta} & 0
\end{array}\right)
\end{eqnarray*}
Squaring, $J_{\; C}^{A}J_{\; B}^{C}=-\delta_{\; B}^{A}$, we see that
$J_{\; B}^{A}$ provides an almost complex structure. That the almost
complex structure is integrable follows immediately in this (global)
basis by the obvious vanishing of the Nijenhuis tensor,
\[
N_{\; BC}^{A}=J_{\; C}^{D}\partial_{D}J_{\; B}^{A}-J_{\; C}^{D}\partial_{D}J_{\; B}^{A}-J_{\; D}^{A}\left(\partial_{C}J_{\; B}^{D}-\partial_{B}J_{\; C}^{D}\right)=0
\]

Next, using the symplectic form to define the compatible metric
\begin{eqnarray*}
g\left(u,v\right) & \equiv & \Omega\left(u,Jv\right)
\end{eqnarray*}
we find that in this basis $g=\left(\begin{array}{cc}
\delta_{\alpha\beta} & 0\\
0 & \delta^{\alpha\beta}
\end{array}\right)$, and we check the remaining compatibility conditions of the triple
$\left(g,J,\Omega\right)$, 
\begin{eqnarray*}
\omega\left(u,v\right) & = & g\left(Ju,v\right)\\
J\left(u\right) & = & \left(\phi_{g}\right)^{-1}\left(\phi_{\omega}\left(u\right)\right)
\end{eqnarray*}
where $\phi_{\omega}$ and $\phi_{g}$ are defined by
\begin{eqnarray*}
\phi_{\omega}\left(u\right) & = & \omega\left(u,\cdot\right)\\
\phi_{g}\left(u\right) & = & g\left(u,\cdot\right)
\end{eqnarray*}
These are easily checked to be satisified, showing that that $\mathcal{M}_{0}^{\left(2n\right)}$
is a Kähler manifold. Notice, however, that the metric of the Kähler
manifold is \emph{not} the restricted Killing metric which we use
in the following considerations.

Finally, a surprising result emerges if we require $\mathcal{M}_{0}^{\left(2n\right)}$
to match our usual expectations for a relativistic phase space. To
make the connection to phase space clear, the precise requirements
were studied in \cite{Spencer:2008p167}, where it was shown that
the flat biconformal gauging of $SO\left(p,q\right)$ in any dimension
$n=p+q$ will have\emph{ Lagrangian submanifolds that are orthogonal
with respect to the 2n-dim biconformal (Killing) metric and have non-degenerate
$n$-dim restrictions of the metric} only if the original space is
Euclidean or signature zero $\left(p\in\left\{ 0,\frac{n}{2},n\right\} \right)$,
and then the signature of the submanifolds is severely limited $\left(p\rightarrow p\pm1\right)$,
leading in the two Euclidean cases to Lorentzian configuration space,
and hence the origin of time. For the case of flat, $8$-dim biconformal
space \cite{Spencer:2008p167} proves the following theorem:

\emph{Flat 8-dim biconformal space is a metric phase space with Lagrangian
submanifolds that are orthogonal with respect to the 2n-dim biconformal
(Killing) metric and have non-degenerate $n$-dim metric restrictions
of the biconformal metric if and only if the initial 4-dim space we
gauge is Euclidean or signature zero. In either of these cases the
resulting configuration sub-manifold is necessarily Lorentzian \cite{Spencer:2008p167}.}

Thus, it is possible to impose the conditions necessary to make biconformal
space a metric phase space only in a restricted subclass of cases,
and the configuration space metric \emph{must} be Lorentzian. In \cite{Spencer:2008p167},
it was found that with a suitable choice of gauge, the metric may
be written in coordinates $y_{\alpha}$ as
\begin{equation}
h_{\alpha\beta}=\frac{1}{\left(y^{2}\right)^{2}}\left(2y_{\alpha}y_{\beta}-y^{2}\delta_{\alpha\beta}\right)\label{eq: FlatBCS_Metric}
\end{equation}
where the signature changing character of the metric is easily seen.

In the metric above, eq.(\ref{eq: FlatBCS_Metric}), $y_{\alpha}=W_{\alpha}$
is the Weyl vector of the space. This points to another unique characteristic
of flat biconformal space. The structures of the conformal group,
treated as described above, give rise to a natural \emph{direction}
of time, given by the gauge field of dilatations. The situation is
reminiscent of previous studies. In 1979, Stelle and West introduced
a special vector field to choose the local symmetry of the MacDowell-Mansouri
theory. The vector breaks the de Sitter symmetry, eliminating the
need for the Wigner-Inönu contraction. Recently, Westman and Zlosnik\cite{Westman:2013mf}
have looked in depth at both the de Sitter and anti-de Sitter cases
using a class of actions which extend that of Stelle and West by including
derivative terms for the vector field and therefore lead to dynamical
symmetry breaking. In \cite{BarberoG.:2003qm,BarberoG.:1995ud} and
Einstein-Aether theory \cite{Jacobson:2008aj}, there is also a special
vector field introduced into the action by hand that can make the
Lorentzian metric Euclidean. These approaches are distinct from that
of the biconformal approach, where the vector necessary for specifying
the timelike direction occurs \emph{naturally} from the underlying
group structure. We will have more to say about this below, where
we show explicitly that the Euclidean gauge theory necessarily posesses
a special \emph{vector}, $\mathbf{v}=\boldsymbol{\omega}-\frac{1}{2}\eta_{ab}\mathbf{d}\eta^{bc}$.
This vector gives the time direction on two Lagrangian submanifolds,
making them necessarily Lorentzian. The full manifold retains its
original symmetry.

\section{A brief note on gravitation\label{sec: gravity}}

Notice that our development to this point was based solely on group
quotients and generalization of the resulting principal fiber bundle.
We have arrived at the form of the curvatures in terms of the Cartan
connection, and Bianchi identities required for integrability, thereby
describing certain classes of geometry. Within the biconformal quotient,
the demand for \emph{orthogonal Lagrangian submanifolds with non-degenerate
$n$-dim restrictions of the Killing metric} leads to the selection
of certain Lorentzian submanifolds. Though our present concern has
to do with the geometric background rather than with gravitational
theories on those backgrounds, for completeness we briefly digress
to specify the action functionals for gravity. The main results of
the current study, taken up again in the final three Sections, concern
only the homogeneous space, $\mathcal{M}_{0}^{\left(2n\right)}$.

We are guided in the choice of action functionals by the example of
general relativity. Given the Riemannin geometries of Section 2.3,
we may write the Einstein-Hilbert action and proceed. More systematically,
however, we may write the most general, even-parity action linear
in the curvature and torsion. This still turns out to be the Einstein-Hilbert
action, and, as noted above, one of the classical field equations
under a full variation of the connection $\left(\delta\mathbf{e}^{b},\delta\boldsymbol{\omega}_{\; b}^{a}\right)$,
implies vanishing torsion. The latter, more robust approach is what
we follow for conformal gravity theories.

It is generally of interest to build the simplest class of actions
possible, and we use the following criteria:
\begin{enumerate}
\item The pure-gravity action should be built from the available curvature
tensor(s) and other tensors which occur in the geometric construction.
\item The action should be of lowest possible order $\geq1$ in the curvatures.
\item The action should be of even parity.
\end{enumerate}
These are of sufficient generality not to bias our choice. It may
also be a reasonable assumption to set certain tensor fields, for
example, the spacetime torsion to zero. This can significantly change
the available tensors, allowing a wider range of action functionals.

Notice that if we perform an infinitesmal \emph{conformal} transformation
to the curvatures, $\left(\boldsymbol{\Omega}_{\;\beta}^{\alpha},\boldsymbol{\Omega}^{\alpha},\boldsymbol{\Omega}_{\beta},\boldsymbol{\Omega}\right)$,
they all mix with one another, since the conformal curvature is really
a single Lie-algebra-valued two form. However, the generalization
to a curved manifold breaks the non-vertical symmetries, allowing
these different components to become independent tensors under the
remaining Weyl group. Thus, to find the available tensors, we apply
an infinitesmal transformation of the \emph{fiber symmetry}. Tensors
are those objects which transform linearly and homogeneously under
these transformations.

\subsection{The auxiliary gauging and Weyl gravity}

The generalization of the auxiliary quotient, $\mathcal{C}/\mathcal{I}\mathcal{W}$,
breaks translational symmetry, and a local transformation of the connection
components immediately shows that the solder form, $\mathbf{e}^{\alpha}$,
becomes a tensor. Correspondingly, the torsion, $\mathbf{T}^{\alpha}$,
no longer mixes with the other curvature components. This suggests
the possibility of a teleparallel theory based on the torsion, but
this would involve little of the conformal structure. Instead we choose
to set $\mathbf{T}^{\alpha}=0$ as an additional condition on our
model. This gives us Riemannian or Weyl geometries instead of Cartan
geometries and is therefore more in line with the requirements of
general relativity.

When the torsion is maintained at zero, both the rotational curvature,
$\boldsymbol{\Omega}_{\;\beta}^{\alpha}$, and the dilatational curvature,
$\boldsymbol{\Omega}$, become tensorial. Because the $n$-dim volume
form has conformal weight $n$ there is no curvature-linear action.
Together with the orthonormal metric and the Levi-Civita tensor, we
build the most general even parity curvature-quadratic action,
\[
S=\int\left(\alpha\boldsymbol{\Omega}_{\;\beta}^{\alpha}\wedge\,^{*}\boldsymbol{\Omega}_{\;\alpha}^{\beta}+\beta\boldsymbol{\Omega}\wedge\,^{*}\boldsymbol{\Omega}\right)
\]
This was partially studied in the 1970s with an eye to supersymmetry
\cite{CrispimRomao:1977hj,CrispimRomao:1978vf,Kaku:1977pa,Kaku:1977rk,Kaku:1978nz},
where the $\beta=0$ case is shown to lead to Weyl gravity. Indeed,
assuming a suitable metric dependence of the remaining connection
components, $\left(\boldsymbol{\omega}_{\;\beta}^{\alpha},\mathbf{f}_{\alpha},\boldsymbol{\omega}\right)$,
metric variation leads to the fourth-order Bach equation \cite{Bach1920}.
However, it has recently been shown that varying all connection forms
independently leads to scale-invariant general relativity \cite{Wheeler:2013ora}.

In dimensions higher than four, our criteria lead to still higher
order actions. Alternatively, curvature-linear actions can be written
in any dimension by introducing a suitable power of a scalar field
\cite{Dirac1936a,Dirac1973a}. This latter reference, \cite{Dirac1973a},
gives the $\phi^{2}R$ action often used in tractor studies.

\subsection{Gravity in the biconformal gauging}

The biconformal gauging, based on $\mathcal{C}/\mathcal{W}$, also
has tensorial basis forms $\left(\boldsymbol{\omega}^{\alpha},\boldsymbol{\omega}_{\alpha}\right)$.
Moreover, each of the component curvatures, $\left(\boldsymbol{\Omega}_{\;\beta}^{\alpha},\boldsymbol{\Omega}^{\alpha},\boldsymbol{\Omega}_{\beta},\boldsymbol{\Omega}\right)$,
becomes an independent tensor under the Weyl group.

In the biconformal case, the volume form $e_{\qquad\alpha\beta\ldots\nu}^{\rho\sigma\ldots\lambda}\boldsymbol{\omega}^{\alpha}\wedge\boldsymbol{\omega}^{\beta}\wedge\ldots\wedge\boldsymbol{\omega}^{\nu}\wedge\boldsymbol{\omega}_{\rho}\wedge\boldsymbol{\omega}_{\sigma}\wedge\ldots\wedge\boldsymbol{\omega}_{\lambda}$
has zero conformal weight. Since both $\boldsymbol{\Omega}_{\;\beta}^{\alpha}$
and $\boldsymbol{\Omega}$ also have zero conformal weight, there
exists a curvature-linear action in any dimension \cite{Wehner:1999p1653}.
The most general linear case is
\[
S=\int\left(\alpha\boldsymbol{\Omega}_{\;\beta}^{\alpha}+\beta\boldsymbol{\Omega}\delta_{\beta}^{\alpha}+\gamma\boldsymbol{\omega}^{\alpha}\wedge\boldsymbol{\omega}_{\beta}\right)\wedge e_{\qquad\alpha\mu\ldots\nu}^{\beta\rho\ldots\sigma}\boldsymbol{\omega}^{\mu}\wedge\ldots\wedge\boldsymbol{\omega}^{\nu}\wedge\boldsymbol{\omega}_{\rho}\wedge\ldots\wedge\boldsymbol{\omega}_{\sigma}
\]
Notice that we now have three important properties of biconformal
gravity that arise because of the doubled dimension: (1) the non-degenerate
conformal Killing metric induces a non-degenerate metric on the manifold,
(2) the dilatational structure equation generically gives a symplectic
form, and (3) there exists a Weyl symmetric action functional linear
in the curvature, valid in any dimension.

There are a number of known results following from the linear action.
In \cite{Wehner:1999p1653} torsion-constrained solutions are found
which are consistent with scale-invariant general relativity. Subsequent
work along the same lines shows that the torsion-free solutions are
determined by the spacetime solder form, and reduce to describe spaces
conformal to Ricci-flat spacetimes on the corresponding spacetime
submanifold \cite{Wheeler:2002}. A supersymmetric version is presented
in \cite{Anderson:2003db}, and studies of Hamiltonian dynamics \cite{Anderson:2004zy,Wheeler:2003bi}
and quantum dynamics \cite{Anderson:2004p612} support the idea that
the models describe some type of relativistic phase space determined
by the configuration space solution.

\section{Homogeneous biconformal space in a conformally orthonormal, symplectic
basis\label{sec:PropertiesHomogeneousBCS}}

The central goal of the remainder of this manuscript is to examine
properties of the homogeneous manifold, $\mathcal{M}_{0}^{\left(2n\right)}$,
which become evident in a conformally orthonormal basis, that is,
a basis which is orthonormal up to an overall conformal factor. Generically,
the properties we discuss will be inherited by the related gravity
theories as well.

As noted above, biconformal space is immediately seen to possess several
structures not seen in other gravitational gauge theories: a non-degenerate
restriction of the Killing metric%
\footnote{There are non-degenerate restrictions in anti-de Sitter and de Sitter
gravitational gauge theories.%
}, a symplectic form, and Kähler structure. In addition, the signature
theorem in \cite{Spencer:2008p167} shows that if the original space
has signature $\pm n$ or zero, the \emph{imposition of involution
conditions} leads to orthogonal Lagrangian submanifolds that have
non-degenerate $n$-dim restrictions of the Killing metric. Further,
\emph{constraining the momentum space} to be as flat as permitted
requires the restricted metrics to be Lorentzian. We strengthen these
results in this Section and the next. Concerning ourselves only with
elements of the \emph{geometry} of the Euclidean $\left(s=\pm n\right)$
cases (as opposed to the additional restrictions of the field equations,
involution conditions or other constraints), we show the presence
of exactly such Lorentzian signature Lagrangian submanifolds \emph{without
further assumptions}.

We go on to study the transformation of the spin connection when we
transform the basis of an $8$-dim biconformal space to one adapted
to the Lagrangian submanifolds. We show that in addition to the Lorentzian
metric, a Lorentzian connection emerges on the configuration and momentum
spaces and there are two new tensor fields. Finally, we examine the
curvature of these Lorentzian connections and find both a cosmological
constant and cosmological ``dust''. While it is premature to make
quantitative predictions, these new geometric features provide novel
candidates for dark energy and dark matter.

\subsection{The biconformal quotient}

We start with the biconformal gauging of Section \ref{sec:Conformal-Quotients},
specialized to the case of compactified, \emph{Euclidean} $\mathbb{R}^{4}$
in a conformally orthonormal, symplectic basis. The Maurer-Cartan
structure equations are

\begin{eqnarray}
\mathbf{d}\boldsymbol{\omega}_{\;\beta}^{\alpha} & = & \boldsymbol{\omega}_{\;\beta}^{\mu}\wedge\boldsymbol{\omega}_{\;\mu}^{\alpha}+2\Delta_{\nu\beta}^{\alpha\mu}\boldsymbol{\omega}_{\mu}\wedge\boldsymbol{\omega}^{\nu}\label{eq:OrigStrucSpinConn}\\
\mathbf{d}\boldsymbol{\omega}^{\alpha} & = & \boldsymbol{\omega}^{\beta}\wedge\boldsymbol{\omega}_{\;\beta}^{\alpha}+\boldsymbol{\omega}\wedge\boldsymbol{\omega}^{\alpha}\label{eq:OrigStrucSolder}\\
\mathbf{d}\boldsymbol{\omega}_{\alpha} & = & \boldsymbol{\omega}_{\;\alpha}^{\beta}\wedge\boldsymbol{\omega}_{\beta}+\boldsymbol{\omega}_{\alpha}\wedge\boldsymbol{\omega}\label{eq:OrigStrucCoSolder}\\
\mathbf{d}\boldsymbol{\omega} & = & \boldsymbol{\omega}^{\alpha}\wedge\boldsymbol{\omega}_{\alpha}\label{eq:OrigStrucWeyl}
\end{eqnarray}
where the connection one-forms represent $SO\left(4\right)$ rotations,
translations, special conformal transformations and dilatations respectively.
The projection operator $\Delta_{\gamma\beta}^{\alpha\mu}\equiv\frac{1}{2}\left(\delta_{\gamma}^{\alpha}\delta_{\beta}^{\mu}-\delta^{\alpha\mu}\delta_{\gamma\beta}\right)$
in eq.(\ref{eq:OrigStrucSpinConn}) gives that part of any $\left({1\atop 1}\right)$-tensor
antisymmetric with respect to the original Euclidean metric, $\delta_{\alpha\beta}$.
As discussed in Section \ref{sub:The-biconformal-gauging}, this group
has a non-degenerate, $15$-dim Killing metric. We stress that the
structure equations and Killing metric \textendash{} and hence their
restrictions to the quotient manifold \textendash{} are intrinsic
to the conformal symmetry.

The gauging begins with the quotient of this conformal group, $SO\left(5,1\right)$,
by its Weyl subgroup, spanned by the connection forms $\boldsymbol{\omega}_{\;\beta}^{\alpha}$
(here dual to $SO(4)$ generators) and $\boldsymbol{\omega}$. The
co-tangent space of the quotient manifold is then spanned by the solder
form, $\boldsymbol{\omega}^{\alpha}$, and the co-solder form, $\boldsymbol{\omega}_{\alpha}$,
and the full conformal group becomes a principal fiber bundle with
local Weyl symmetry over this $8$-dim quotient manifold. The independence
of $\boldsymbol{\omega}^{\alpha}$ and $\boldsymbol{\omega}_{\alpha}$
in the biconformal gauging makes the $2$-form $\boldsymbol{\omega}^{\alpha}\wedge\boldsymbol{\omega}_{\alpha}$
non-degenerate, and eq.(\ref{eq:OrigStrucWeyl}) immediately shows
that $\boldsymbol{\omega}^{\alpha}\wedge\boldsymbol{\omega}_{\alpha}$
is a symplectic form. The basis $\left(\boldsymbol{\omega}^{\alpha},\boldsymbol{\omega}_{\alpha}\right)$
is canonical.

The involution evident in eq.(\ref{eq:OrigStrucSolder}) shows that
the solder forms, $\boldsymbol{\omega}^{\alpha}$, span a submanifold,
and from the simultaneous vanishing of the symplectic form this submanifold
is Lagrangian. Similarly, eq.(\ref{eq:OrigStrucCoSolder}) shows that
the $\boldsymbol{\omega}_{\beta}$ span a Lagrangian submanifold.
However, notice that neither of these submanifolds, spanned by either
$\boldsymbol{\omega}^{\alpha}$ or $\boldsymbol{\omega}_{\alpha}$
alone, has an induced metric, since by eq.\ref{eq:KillingMetric-1},
$\left\langle \boldsymbol{\omega}^{\alpha},\boldsymbol{\omega}^{\beta}\right\rangle =\left\langle \boldsymbol{\omega}_{\alpha},\boldsymbol{\omega}_{\beta}\right\rangle =0$.
The \emph{orthonormal} basis will make the Killing metric block diagonal,
guaranteeing that its restriction to the configuration and momentum
submanifolds have well-defined, non-degenerate metrics.

It was shown in \cite{Spencer:2008p167} this it is consistent (for
signatures $\pm n,\,0$ \emph{only}) to impose involution conditions
and momentum flatness in this rotated basis in such a way that the
new basis still gives Lagrangian submanifolds. Moreover, the restriction
of the Killing metric to these new submanifolds is necessarily Lorentzian.
In what follows, we do not need the assumptions of momentum flatness
or involution, and work only with intrinsic properties of $\mathcal{M}_{0}^{\left(2n\right)}$.
This Section describes the new basis and resulting connection, while
the next establishes that for initial Euclidean signature, the principal
results of \cite{Spencer:2008p167} follow necessarily. Our results
show that the timelike directions in these models arise from intrinsically
conformal structures.

We now change to a new canonical basis, adapted to the Lagrangian
submanifolds.

\subsection{The conformally-orthonormal Lagrangian basis}

In \cite{Spencer:2008p167} the $\left(\boldsymbol{\omega}^{\alpha},\boldsymbol{\omega}_{\alpha}\right)$
basis is rotated so that the metric, $h_{AB}$ becomes block diagonal
\begin{eqnarray*}
\left[\begin{array}{cc}
0 & \delta_{\beta}^{\alpha}\\
\delta_{\beta}^{\alpha} & 0
\end{array}\right]\Rightarrow\left[h_{AB}\right] & = & \left[\begin{array}{cc}
h_{ab} & 0\\
0 & -h^{ab}
\end{array}\right]
\end{eqnarray*}
while the symplectic form remains canonical. This makes the Lagrangian
submanifolds orthogonal with a non-degenerate restriction to the metric.
Here we use the same basis change, but in addition define coefficients,
$h_{a}^{\;\alpha}$ to relate the orthogonal metric to one conformally
orthonormal on the submanifolds, $\eta_{ab}=h_{a}^{\;\alpha}h_{\alpha\beta}h_{b}^{\;\beta}$,
where $\eta_{ab}$ is conformal to $diag\left(\pm1,\pm1,\pm1,\pm1\right)$.
From \cite{Spencer:2008p167} we know that $h_{ab}$ is necessarily
Lorentzian, $h_{ab}=\eta_{ab}=e^{2\phi}diag\left(-1,1,1,1\right)=e^{2\phi}\eta_{ab}^{0}$
and we give a more general proof below. Notice that the definition
of $\eta_{ab}$ includes an unknown conformal factor. The required
change of basis is then 
\begin{eqnarray}
\mathbf{e}^{a} & = & h_{\alpha}^{\; a}\left(\boldsymbol{\omega}^{\alpha}+\frac{1}{2}h^{\alpha\beta}\boldsymbol{\omega}_{\beta}\right)\label{eq:NewInTermsOfOld_e}\\
\mathbf{f}_{a} & = & h_{a}^{\;\alpha}\left(\frac{1}{2}\boldsymbol{\omega}_{\alpha}-h_{\alpha\beta}\boldsymbol{\omega}^{\beta}\right)\label{eq:NewInTermsOfOld_f}
\end{eqnarray}
with inverse basis change
\begin{eqnarray}
\boldsymbol{\omega}^{\alpha} & = & \frac{1}{2}h_{a}^{\;\alpha}\left(\mathbf{e}^{a}-\eta^{ab}\mathbf{f}_{b}\right)\label{eq:OldInTermsOfNew1-1}\\
\boldsymbol{\omega}_{\alpha} & = & h_{\alpha}^{\; a}\left(\mathbf{f}_{a}+\eta_{ab}\mathbf{e}^{b}\right)\label{eq:OldInTermsOfNew2-1}
\end{eqnarray}
Using (\ref{eq:KillingMetric-1}), the Killing metric is easily checked
to be
\begin{eqnarray*}
\left[\begin{array}{cc}
\left\langle \mathbf{e}^{a},\mathbf{e}^{b}\right\rangle  & \left\langle \mathbf{e}^{a},\mathbf{f}_{b}\right\rangle \\
\left\langle \mathbf{f}_{a},\mathbf{e}^{b}\right\rangle  & \left\langle \mathbf{f}_{a},\mathbf{f}_{b}\right\rangle 
\end{array}\right] & = & \left[\begin{array}{cc}
h_{\alpha}^{\; a}h_{\beta}^{\; b}h^{\left(\alpha\beta\right)} & 0\\
0 & -h_{a}^{\;\alpha}h_{b}^{\;\beta}h_{\left(\alpha\beta\right)}
\end{array}\right]\\
 & = & \left[\begin{array}{cc}
e^{-2\phi}\eta_{0}^{ab} & 0\\
0 & -e^{2\phi}\eta_{ab}^{0}
\end{array}\right]
\end{eqnarray*}
where $h_{\alpha\beta}=h_{\left(\alpha\beta\right)}$, and $h^{\alpha\beta}h_{\beta\gamma}=\delta_{\gamma}^{\alpha}$.

The new basis is also canonical, as we see by transforming the dilatation
equation, eq.(\ref{eq:OrigStrucWeyl}), to find $\mathbf{d}\boldsymbol{\omega}=\mathbf{e}^{a}\mathbf{f}_{a}$.
We refer to the $\mathbf{f}_{a}=0$ and $\mathbf{e}^{a}=0$ submanifolds
as the configuration and momentum submanifolds respectively.

\subsection{Properties of the structure equations in the new basis\label{sub:OrthonormalProperties}}

We now explore the properties of the biconformal system in this adapted
basis. Rewriting the remaining structure equations, eqs.(\ref{eq:OrigStrucSpinConn},
\ref{eq:OrigStrucSolder}, \ref{eq:OrigStrucCoSolder}), in terms
of $\mathbf{e}^{a}$ and $\mathbf{f}_{a}$, we show some striking
cancelations that lead to the emergence of a connection compatible
with the Lorentzian metric, and two new tensors.

We begin with the exterior derivative of eq.(\ref{eq:NewInTermsOfOld_e}),
using structure equations eq.(\ref{eq:OrigStrucSolder}) and eq.(\ref{eq:OrigStrucCoSolder}),
and then using the basis change equations eqs.(\ref{eq:OldInTermsOfNew1-1},
\ref{eq:OldInTermsOfNew2-1}). Because eqs.(\ref{eq:OldInTermsOfNew1-1},
\ref{eq:OldInTermsOfNew2-1}) involve the sum and difference of $\mathbf{e}^{a}$
and $\mathbf{f}_{b}$, separating by these new basis forms leads to
a separation of symmetries. This leads to a cumbersome expansion,
which reduces considerably and in significant ways, to 
\begin{eqnarray}
\mathbf{d}\mathbf{e}^{a} & = & \mathbf{e}^{b}\wedge\Theta_{cb}^{ad}\boldsymbol{\tau}_{\; d}^{c}-\eta^{bc}\mathbf{f}_{c}\wedge\Xi_{db}^{ae}\boldsymbol{\tau}_{\; e}^{d}+\frac{1}{2}\eta_{bc}\mathbf{d}\eta^{ab}\wedge\mathbf{e}^{c}+\frac{1}{2}\mathbf{d}\eta^{ab}\wedge\mathbf{f}_{b}+2\eta^{ab}\mathbf{f}_{b}\wedge\boldsymbol{\omega}\label{ugly transformation}
\end{eqnarray}
where we define projections $\Theta_{db}^{ac}\equiv\frac{1}{2}\left(\delta_{d}^{a}\delta_{b}^{c}-\eta^{ac}\eta_{bd}\right)$
and $\Xi_{cb}^{ad}\equiv\frac{1}{2}\left(\delta_{c}^{a}\delta_{b}^{d}+\eta^{ad}\eta_{cb}\right)$
that separate symmetries with respect to the new metric $\eta_{ab}$
rather than $\delta_{\alpha\beta}$. These give the antisymmetric
and symmetric parts, respectively, of a $\left({1\atop 1}\right)$-tensor
with respect to the \emph{new} orthonormal metric, $\eta_{ab}$. Notice
that these projections are independent of the conformal factor on
$\eta_{ab}$.

The significance of the reduction lies in how the symmetries separate
between the different subspaces. Just as the curvatures split into
three parts, eq.(\ref{ugly transformation}) and each of the remaining
structure equations splits into three parts. Expanding these independent
parts separately allows us to see the Riemannian structure of the
configuration and momentum spaces. It is useful to first define
\begin{eqnarray}
\boldsymbol{\tau}_{\; b}^{a} & \equiv & \boldsymbol{\alpha}_{\; b}^{a}+\boldsymbol{\beta}_{\; b}^{a}\label{eq:SpinConnProjDef-1-1}
\end{eqnarray}
where $\boldsymbol{\alpha}_{\; b}^{a}\equiv\Theta_{cb}^{ad}\boldsymbol{\tau}_{\; d}^{c}$
and $\boldsymbol{\beta}_{\; b}^{a}\equiv\Xi_{cb}^{ad}\boldsymbol{\tau}_{\; d}^{c}$.
Then, to facilitate the split into $\mathbf{e}^{a}\wedge\mathbf{e}^{b}$,
$\mathbf{e}^{a}\wedge\mathbf{f}_{b}$ and $\mathbf{f}_{a}\wedge\mathbf{f}_{b}$
parts, we partition the spin connection and Weyl vector by submanifold,
defining 
\begin{eqnarray}
\boldsymbol{\alpha}_{\; b}^{a} & \equiv & \boldsymbol{\sigma}_{\; b}^{a}+\boldsymbol{\gamma}_{\; b}^{a}=\sigma_{\; bc}^{a}\mathbf{e}^{c}+\gamma_{\; b}^{a\; c}\mathbf{f}_{c}\label{eq:antisymmetricSpin Connection}\\
\boldsymbol{\beta}_{\; b}^{a} & \equiv & \boldsymbol{\mu}_{\; b}^{a}+\boldsymbol{\rho}_{\; b}^{a}=\mu_{\; bc}^{a}\mathbf{e}^{c}+\rho_{\; b}^{a\; c}\mathbf{f}_{c}\label{eq:symmetricSpin Connection}\\
\boldsymbol{\omega} & \equiv & W_{a}\mathbf{e}^{a}+W^{a}\mathbf{f}_{a}\label{eq:WeylVectorDecomp}
\end{eqnarray}
We also split the exterior derivative, $\mathbf{d}=\mathbf{d}_{\left(x\right)}+\mathbf{d}_{\left(y\right)}$,
where coordinates $x^{\alpha}$ and $y_{\alpha}$ are used on the
$\mathbf{e}^{a}=e_{\alpha}^{\quad a}\mathbf{d}x^{\alpha}$ and $\mathbf{f}_{a}=f_{a}^{\quad\alpha}\mathbf{d}y_{\alpha}$
submanifolds, respectively. Using these, we expand each of the structure
equations into three $\mathcal{W}$-invariant parts. The complete
set (with curvatures included for completeness) is given in Appendix
1.

The simplifying features and notable properties include: 
\begin{enumerate}
\item \emph{The new connection: }The first thing that is evident is that
all occurences of the spin connection $\boldsymbol{\omega}_{\;\beta}^{\alpha}$
may be written in terms of the combination
\begin{equation}
\boldsymbol{\tau}_{\; b}^{a}\equiv h_{\alpha}^{\; a}\boldsymbol{\omega}_{\;\beta}^{\alpha}h_{b}^{\;\beta}-h_{b}^{\;\alpha}\mathbf{d}h_{\alpha}^{\; a}\label{eq:DefinitionTau}
\end{equation}
which, as we show below, transforms as a Lorentz spin connection.
Although the basis change is not a gauge transformation, the change
in the connection has a similar inhomogeneous form. Because $h_{\alpha}^{\; a}$
is a change of basis rather than local $SO\left(n\right)$ or local
Lorentz, the inhomogeneous term has no particular symmetry property,
so $\boldsymbol{\tau}_{\; b}^{a}$ will have both symmetric and antisymmetric
parts.
\item \emph{Separation of symmetric and antisymmetric parts:} Notice in
eq.(\ref{ugly transformation}) how the antisymmetric part of the
new connection, $\boldsymbol{\alpha}_{\; b}^{a}$, is associated with
$\mathbf{e}^{b}$, while the symmetric part, $\boldsymbol{\beta}_{\; b}^{a}$
pairs with $\mathbf{f}_{c}$. This surprising correspondence puts
the symmetric part into the cross-terms while leaving the connection
of the configuration submanifold metric compatible, up to the conformal
factor. 
\item \emph{Cancellation of the submanifold Weyl vector:} The Weyl vector
terms cancel on the configuration submanifold, while the $\mathbf{f}_{a}$
terms add. The expansion of the $\mathbf{d}\mathbf{f}_{a}$ structure
equation shows that the Weyl vector also drops out of the momentum
submanifold equations. Nonetheless, these submanifold equations are
scale invariant because of the residual metric derivative. Recognizing
the combination of $\mathbf{d}h$ terms that arises as $\mathbf{d}\eta^{ab}$,
and recalling that $\eta_{ab}=e^{2\phi}\eta_{ab}^{0}$, we have $-\frac{1}{2}\mathbf{d}\eta^{ac}\eta_{cb}=\delta_{b}^{a}\mathbf{d}\phi$.
When the metric is rescaled, this term changes with the same inhomogeneous
term as the Weyl vector.
\item \emph{Covariant derivative and a second Weyl-type connection:} It
is natural to define the $\boldsymbol{\tau}_{\; c}^{b}$-covariant
derivative of the metric. Since $\eta^{cb}\boldsymbol{\alpha}_{\; c}^{a}+\eta^{ac}\boldsymbol{\alpha}_{\; c}^{b}=0$,
it depends only on $\boldsymbol{\beta}_{\; c}^{a}$ and the Weyl vector,
\begin{eqnarray}
\mathbf{D}\eta^{ab} & \equiv & \mathbf{d}\eta^{ab}+\eta^{cb}\boldsymbol{\tau}_{\; c}^{a}+\eta^{ac}\boldsymbol{\tau}_{\; c}^{b}-2\boldsymbol{\omega}\eta^{ae}\label{eq:Derivative of metric}\\
 & = & \mathbf{d}\eta^{ab}+2\eta^{cb}\boldsymbol{\beta}_{\; c}^{a}-2\boldsymbol{\omega}\eta^{ab}\label{eq:Reduced derivative}
\end{eqnarray}
This derivative allows us to express the structure of the biconformal
space in terms of the Lorentzian properties.
\end{enumerate}
When all of the identifications and definitions are included, and
carrying out similar calculations for the remaining structure equations,
the full set becomes
\begin{eqnarray}
\mathbf{d}\boldsymbol{\tau}_{\; b}^{a} & = & \boldsymbol{\tau}_{\; b}^{c}\wedge\boldsymbol{\mathbf{\tau}}_{\; c}^{a}+\Delta_{db}^{ae}\eta_{ec}\mathbf{e}^{c}\wedge\mathbf{e}^{d}-\Delta_{eb}^{ac}\eta^{ed}\mathbf{f}_{c}\wedge\mathbf{f}_{d}+2\Delta_{fb}^{ae}\Xi_{de}^{fc}\mathbf{f}_{c}\wedge\mathbf{e}^{d}\label{eq:NewStrucSpinConn}\\
\mathbf{d}\mathbf{e}^{a} & = & \mathbf{e}^{c}\wedge\boldsymbol{\alpha}_{\; c}^{a}+\frac{1}{2}\eta_{cb}\mathbf{d}\eta^{ac}\wedge\mathbf{e}^{b}+\frac{1}{2}\mathbf{D}\eta^{ab}\wedge\mathbf{f}_{b}\label{eq:NewStrucSolder}\\
\mathbf{d}\mathbf{f}_{a} & = & \boldsymbol{\alpha}_{\; a}^{b}\wedge\mathbf{f}_{b}+\frac{1}{2}\eta^{bc}\mathbf{d}\eta_{ab}\wedge\mathbf{f}_{c}-\frac{1}{2}\mathbf{D}\eta_{ab}\wedge\mathbf{e}^{b}\label{eq:NewStrucCoSolder}\\
\mathbf{d}\boldsymbol{\omega} & = & \mathbf{e}^{a}\wedge\mathbf{f}_{a}\label{eq:NewStrucWeyl}
\end{eqnarray}
with the complete $\mathcal{W}$-invariant separation in Appendix
1.

\subsection{Gauge transformations and new tensors}

The biconformal bundle now allows local Lorentz transformations and
local dilatations on $\mathcal{M}_{0}^{\left(2n\right)}$. Under local
Lorentz transformations, $\Lambda_{\; c}^{a}$, the connection $\boldsymbol{\tau}_{\; b}^{a}$
changes with an inhomogeneous term of the form $\bar{\Lambda}_{\; b}^{c}\mathbf{d}\Lambda_{\; c}^{a}$.
Since this term lies in the Lie algebra of the Lorentz group, it is
antisymmetric with respect to $\eta_{ab}$, $\Theta_{db}^{ac}\left(\bar{\Lambda}_{\; c}^{e}\mathbf{d}\Lambda_{\; e}^{d}\right)=\bar{\Lambda}_{\; b}^{e}\mathbf{d}\Lambda_{\; e}^{a}$and
therefore only changes the corresponding $\Theta_{db}^{ac}$-antisymmetric
part of the connection, with the symmetric part transforming homogeneously:
\begin{eqnarray*}
\tilde{\boldsymbol{\alpha}}_{\; b}^{a} & = & \Lambda_{\; c}^{a}\boldsymbol{\alpha}_{\; d}^{c}\bar{\Lambda}_{\; b}^{d}-\bar{\Lambda}_{\; b}^{c}\mathbf{d}\Lambda_{\; c}^{a}\\
\tilde{\boldsymbol{\beta}}_{\; b}^{a} & = & \Lambda_{\; c}^{a}\boldsymbol{\beta}_{\; d}^{c}\bar{\Lambda}_{\; b}^{d}
\end{eqnarray*}
Having no inhomogeneous term, $\boldsymbol{\beta}_{\; b}^{a}$ is
a Lorentz tensor. This new tensor field $\boldsymbol{\beta}_{\; b}^{a}$
necessarily includes degrees of freedom from the original connection
that cannot be present in $\boldsymbol{\alpha}_{\; b}^{a}$, the total
equaling the degrees of freedom present in $\boldsymbol{\tau}_{\; b}^{a}$.
As there is no obvious constraint on the connection $\boldsymbol{\alpha}_{\; b}^{a}$,
we expect $\boldsymbol{\beta}_{\; b}^{a}$ to be highly constrained.
Clearly, $\boldsymbol{\alpha}_{\; d}^{c}$ transforms as a Lorentzian
spin connection, and the addition of the tensor $\boldsymbol{\beta}_{\; b}^{a}$
preserves this property, so $\boldsymbol{\tau}_{\; b}^{a}$ is a local
Lorentz connection. 

Transformation of the connection under dilatations reveals another
new tensor. The Weyl vector transforms inhomogeneously in the usual
way, $\tilde{\boldsymbol{\omega}}=\boldsymbol{\omega}+\mathbf{d}f$,
but, as noted above, the expression $\frac{1}{2}\eta_{cb}\mathbf{d}\eta^{ac}$
also transforms,
\[
\frac{1}{2}\tilde{\eta}_{cb}\mathbf{d}\tilde{\eta}^{ac}=\delta_{b}^{a}\mathbf{d}\tilde{\phi}=\delta_{b}^{a}\left(\mathbf{d}\phi-\mathbf{d}f\right)
\]
so that the combination
\[
\mathbf{v}=\boldsymbol{\omega}+\mathbf{d}\phi
\]
is scale invariant. Notice the presence of two distinct scalars here.
Obviously, given $\frac{1}{2}\eta^{ac}\mathbf{d}\eta_{cb}=\delta_{b}^{a}\mathbf{d}\phi$
we can choose a gauge function, $f_{1}=-\phi$, such that $\frac{1}{2}\eta^{ac}\mathbf{d}\eta_{cb}=0$.
We also have, $\mathbf{d}\boldsymbol{\omega}=0,$ on the configuration
submanifold, so that $\boldsymbol{\omega}=\mathbf{d}f_{2}$, for some
scalar $f_{2}$ and this might be gauged to zero instead. But while
one or the other of $\boldsymbol{\omega}$ or $\mathbf{d}\phi$ can
be gauged to zero, their \emph{sum} is gauge invariant. As we show
below, it is the resulting vector $\mathbf{v}$ which determines the
timelike directions.

Recall that certain involution relationships must be satisfied to
ensure that spacetime and momentum space are each submanifolds. The
involution conditions in homogeneous biconformal space are 
\begin{eqnarray}
0 & = & \boldsymbol{\mu}_{\; b}^{a}\wedge\mathbf{e}^{b}-\mathbf{v}_{\left(x\right)}\wedge\mathbf{e}^{a}\label{eq:InvolutionE}\\
0 & = & \boldsymbol{\rho}_{\;\; a}^{b}\wedge\mathbf{f}_{b}-\mathbf{u}_{\left(y\right)}\wedge\mathbf{f}_{a}\label{eq:InvolutionF}
\end{eqnarray}
where $\mathbf{v}\equiv\mathbf{v}_{\left(x\right)}+\mathbf{u}_{\left(y\right)}\equiv v_{a}\mathbf{e}^{a}+u^{a}\mathbf{f}_{a}$.
These were imposed as constraints in \cite{Spencer:2008p167}, but
are shown below to hold automatically in Euclidean cases.

\section{Riemannian spacetime in Euclidean biconformal space\label{sec:RiemannianStrucInBCS}}

The principal result of \cite{Spencer:2008p167} was to show that
the flat biconformal space $\mathcal{M}_{0}^{\left(2n\right)}$ arising
from any $SO\left(p,q\right)$ symmetric biconformal gauging can be
identified with a metric phase space only when the initial $n$-space
is of signature $\pm n$ or zero. To make the identification, involution
of the Lagrangian submanifolds was imposed, and it was assumed that
the momentum space is conformally flat. With these assumptions the
Lagrangian configuration and momentum submanifolds of the signature
$\pm n$ cases are necessarily Lorentzian.

Here we substantially strengthen this result, by considering only
the Euclidean case. We are able to show that further assumptions are
unnecessary. The gauging \emph{always} leads to Lorentzian configuration
and momentum submanifolds, the involution conditions are automatically
satisfied by the structure equations, and both the configuration and
momentum spaces are conformally flat. We make no assumptions beyond
the choice of the quotient $\mathcal{C}/\mathcal{W}$ and the structures
that follow from these groups.

Because this result shows the development of the Lorentzian metric
on the Lagrangian submanifolds, we give details of the calculation.

\subsection{Solution of the structure equations}

A complete solution of the structure equations in the original basis,
eqs.(\ref{eq:OrigStrucSpinConn}-\ref{eq:OrigStrucWeyl}) is given
in \cite{Wheeler:1997pc} and derived in \cite{Wehner:1999p1653},
with a concise derivation presented in \cite{Anderson:2004zy}. By
choosing the gauge and coordinates $\left(w^{\alpha},s_{\beta}\right)$
appropriately, where Greek indices now refer to \emph{coordinates}
and will do so for the remainder of this manuscript\footnote[6]{The connection forms could be written with distinct indices, for example
as $\boldsymbol{\omega}^{a}=\delta_{\alpha}^{a}\mathbf{d}w^{\alpha}$,
but this is unnecessarily cumbersome.%
}, the solution may be given the form,
\begin{eqnarray}
\boldsymbol{\omega}_{\;\beta}^{\alpha} & = & 2\Delta_{\nu\beta}^{\alpha\mu}s_{\mu}\mathbf{d}w^{\nu}\label{eq:original solution spin connection}\\
\boldsymbol{\omega}^{\alpha} & = & \mathbf{d}w^{\alpha}\label{original solution solder form}\\
\boldsymbol{\omega}_{\alpha} & = & \mathbf{d}s_{\alpha}-\left(s_{\alpha}s_{\beta}-\frac{1}{2}s^{2}\delta_{\alpha\beta}\right)\mathbf{d}w^{\beta}\label{original solution co-solder}\\
\boldsymbol{\omega} & = & -s_{\alpha}\mathbf{d}w^{\alpha}\label{original solution weyl vector}
\end{eqnarray}
as is easily checked by direct substitution. Our first goal is to
express this solution in the adapted basis and find the resulting
metric.

From the original form of the Killing metric, eq.(\ref{eq:KillingMetric-1}),
we find
\begin{eqnarray*}
\left[\begin{array}{cc}
\left\langle \mathbf{d}w^{\alpha},\mathbf{d}w^{\beta}\right\rangle  & \left\langle \mathbf{d}w^{\alpha},\mathbf{d}s_{\beta}\right\rangle \\
\left\langle \mathbf{d}s_{\alpha},\mathbf{d}w^{\beta}\right\rangle  & \left\langle \mathbf{d}s_{\alpha},\mathbf{d}s_{\beta}\right\rangle 
\end{array}\right] & = & \left[\begin{array}{cc}
0 & \delta_{\beta}^{\alpha}\\
\delta_{\alpha}^{\beta} & -k_{\alpha\beta}
\end{array}\right]
\end{eqnarray*}
where we define $k_{\alpha\beta}\equiv s^{2}\delta_{\alpha\beta}-2s_{\alpha}s_{\beta}$.
This shows that $\mathbf{d}w^{\alpha}$ and $\mathbf{d}s_{\alpha}$
do not span orthogonal subspaces. We want to find the most general
set of orthogonal Lagrangian submanifolds, and the restriction of
the Killing metric to them.

Suppose we find linear combinations of the orginal basis $\boldsymbol{\kappa}^{\beta},\boldsymbol{\lambda}_{\alpha}$
that make the metric block diagonal, with $\boldsymbol{\lambda}_{\alpha}=0$
and $\boldsymbol{\kappa}^{\beta}=0$ giving Lagrangian submanifolds.
Then any further transformation,
\begin{eqnarray*}
\tilde{\boldsymbol{\kappa}}^{\alpha} & = & A_{\;\beta}^{\alpha}\boldsymbol{\kappa}^{\beta}\\
\tilde{\boldsymbol{\lambda}}_{\alpha} & = & B_{\;\alpha}^{\beta}\boldsymbol{\lambda}_{\beta}
\end{eqnarray*}
leaves these submanifolds unchanged and is therefore equivalent. Now
suppose one of the linear combinations is
\begin{eqnarray*}
\tilde{\boldsymbol{\lambda}}_{\alpha} & = & \alpha A_{\;\alpha}^{\beta}\mathbf{d}s_{\beta}+\beta\tilde{C}_{\alpha\mu}\mathbf{d}w^{\mu}\\
 & = & A_{\;\alpha}^{\beta}\left(\alpha\mathbf{d}s_{\beta}+\beta C_{\beta\mu}\mathbf{d}w^{\mu}\right)
\end{eqnarray*}
where $C=A^{-1}\tilde{C}$ and the constants are required to keep
the transformation nondegenerate. Then $\boldsymbol{\lambda}_{\alpha}=\alpha\mathbf{d}s_{\alpha}+\beta C_{\alpha\beta}\mathbf{d}w^{\beta}$
spans the same subspace. A similar argument holds for $\tilde{\boldsymbol{\kappa}}{}^{\beta}$,
so if we can find a basis at all, there is also one of the form
\begin{eqnarray*}
\boldsymbol{\lambda}_{\alpha} & = & \alpha\mathbf{d}s_{\alpha}+\beta C_{\alpha\beta}\mathbf{d}w^{\beta}\\
\boldsymbol{\kappa}^{\alpha} & = & \mu\mathbf{d}w^{\alpha}+\nu B^{\alpha\beta}\mathbf{d}s_{\beta}
\end{eqnarray*}
Now check the symplectic condition,
\begin{eqnarray*}
\boldsymbol{\kappa}^{\alpha}\boldsymbol{\lambda}_{\alpha} & = & \left(\mu\beta C_{\alpha\mu}\right)\mathbf{d}w^{\alpha}\mathbf{d}w^{\mu}+\alpha\mu\left(\delta_{\mu}^{\beta}-\nu\beta C_{\alpha\mu}B^{\alpha\beta}\right)\mathbf{d}w^{\mu}\mathbf{d}s_{\beta}+\left(\nu\alpha B^{\alpha\beta}\right)\mathbf{d}s_{\beta}\mathbf{d}s_{\alpha}
\end{eqnarray*}
To have $\boldsymbol{\kappa}^{\alpha}\boldsymbol{\lambda}_{\alpha}=\mathbf{d}w^{\alpha}\mathbf{d}s_{\alpha}$,
$B^{\alpha\beta}$ and $C_{\alpha\beta}$ must be symmetric and
\begin{eqnarray*}
B=B^{t} & = & \frac{\alpha\mu-1}{\nu\beta}C^{-1}\equiv\alpha\beta\bar{C}
\end{eqnarray*}
Replacing $B_{\alpha\beta}$ in the basis, we look at orthogonality
of the inner product, requiring
\begin{eqnarray*}
0 & = & \left\langle \boldsymbol{\kappa}^{\alpha},\boldsymbol{\lambda}_{\beta}\right\rangle \\
 & = & \left\langle \mu\mathbf{d}w^{\alpha}+\frac{\alpha\mu-1}{\beta}\bar{C}^{\alpha\mu}\mathbf{d}s_{\mu},\alpha\mathbf{d}s_{\beta}+\beta C_{\beta\nu}\mathbf{d}w^{\nu}\right\rangle \\
 & = & \left(2\alpha\mu-1\right)\delta_{\beta}^{\alpha}-\frac{1}{\beta}\alpha\left(\alpha\mu-1\right)\bar{C}^{\alpha\mu}k_{\mu\beta}
\end{eqnarray*}
with solution $C_{\alpha\beta}=\frac{\alpha\left(\alpha\mu-1\right)}{\beta\left(2\alpha\mu-1\right)}k_{\alpha\beta}$.
Therefore, the basis 
\begin{eqnarray*}
\boldsymbol{\lambda}_{\alpha} & = & \alpha\mathbf{d}s_{\alpha}+\frac{\alpha\left(\alpha\mu-1\right)}{\left(2\alpha\mu-1\right)}k_{\alpha\beta}\mathbf{d}w^{\beta}\\
\boldsymbol{\kappa}^{\alpha} & = & \mu\mathbf{d}w^{\alpha}+\frac{2\alpha\mu-1}{\alpha}k^{\alpha\beta}\mathbf{d}s_{\beta}
\end{eqnarray*}
satisfies the required properties and is equivalent to any other which
does.

The metric restrictions to the submanifolds are now immediate from
the inner products:
\begin{eqnarray*}
\left\langle \boldsymbol{\kappa}^{\alpha},\boldsymbol{\kappa}^{\beta}\right\rangle  & = & \frac{2\alpha\mu-1}{\alpha^{2}}k^{\alpha\beta}\\
\left\langle \boldsymbol{\lambda}_{\alpha},\boldsymbol{\lambda}_{\beta}\right\rangle  & = & -\frac{\alpha^{2}}{2\alpha\mu-1}k_{\alpha\beta}
\end{eqnarray*}
This shows that the metric on the Lagrangian submanifolds is proportional
to $k_{\alpha\beta}$, and we normalize the proportionality to 1 by
choosing $\mu=\frac{1+k\alpha^{2}}{2\alpha}$ and $\beta\equiv k\alpha$,
where $k=\pm1$. This puts the basis in the form 
\begin{eqnarray*}
\boldsymbol{\kappa}^{\alpha} & = & \frac{k}{2\beta}\left(\left(k\beta^{2}+1\right)\mathbf{d}w^{\alpha}+2k\beta^{2}k^{\alpha\beta}\mathbf{d}s_{\beta}\right)\\
\boldsymbol{\lambda}_{\alpha} & = & \frac{1}{2\beta}\left(2k\beta^{2}\mathbf{d}s_{\alpha}+\left(k\beta^{2}-1\right)k_{\alpha\beta}\mathbf{d}w^{\beta}\right)
\end{eqnarray*}

Now that we have established the metric
\[
k_{\alpha\beta}=s^{2}\left(\delta_{\alpha\beta}-\frac{2}{s^{2}}s_{\alpha}s_{\beta}\right)
\]
where $\delta_{\alpha\beta}$ is the Euclidean metric and $s^{2}=\delta^{\alpha\beta}s_{\alpha}s_{\beta}>0$,
and have found one basis for the submanifolds, we may form an orthonormal
basis for each, setting $\eta_{ab}=h_{a}^{\;\alpha}h_{b}^{\;\beta}k_{\alpha\beta}$.
\begin{eqnarray}
\mathbf{e}^{a} & = & \frac{k}{2\beta}h_{\alpha}^{\; a}\left(\left(1+k\beta^{2}\right)\mathbf{d}w^{\alpha}+2k\beta^{2}k^{\alpha\beta}\mathbf{d}s_{\beta}\right)\label{Solution solder form}\\
\mathbf{f}_{a} & = & \frac{1}{2\beta}h_{a}^{\;\alpha}\left(2k\beta^{2}\mathbf{d}s_{\alpha}-\left(1-k\beta^{2}\right)k_{\alpha\beta}\mathbf{d}w^{\beta}\right)\label{Solution co-solder form}
\end{eqnarray}
We see from the form $k_{\alpha\beta}=s^{2}\left(\delta_{\alpha\beta}-\frac{2}{s^{2}}s_{\alpha}s_{\beta}\right)$
that at any point $s_{\alpha}^{0}$, a rotation that takes $\frac{1}{\sqrt{s^{w}}}s_{\alpha}^{0}$
to a fixed direction $\hat{\mathbf{n}}$ will take $k_{\alpha\beta}$
to $s^{2}diag\left(-1,1,\ldots,1\right)$ so the orthonormal metric
$\eta_{ab}$ is Lorentzian. This is one of our central results. Since
eqs.(\ref{eq:original solution spin connection}-\ref{original solution weyl vector})
provide an exact, general solution to the structure equations, the
induced configuration and momentum spaces of Euclidean biconformal
spaces are \emph{always} Lorentzian, without restrictions.

We now find the connections forms in the orthogonal basis and check
the involution conditions required to guarantee that the configuration
and momentum subspaces are Lagrangian submanifolds.

\subsection{The connection in the adapted solution basis}

We have defined $\boldsymbol{\tau}_{\; b}^{a}$ in eq. (\ref{eq:DefinitionTau})
with antisymmetric and symmetric parts $\boldsymbol{\alpha}_{\; b}^{a}$
and $\boldsymbol{\beta}_{\; b}^{a}$, subdivided between the $\mathbf{e}^{a}$
and $\mathbf{f}_{a}$ subspaces, eq(\ref{eq:antisymmetricSpin Connection},
\ref{eq:symmetricSpin Connection}). All quantities may be written
in terms of the new basis. We will make use of $s_{a}\equiv h_{a}^{\;\alpha}s_{\alpha}$
and $\delta_{ab}\equiv h_{a}^{\;\alpha}h_{b}^{\;\beta}\delta_{\alpha\beta}$.
In terms of these, the orthonormal metric is $\eta_{ab}=s^{2}\left(\delta_{ab}-\frac{2}{s^{2}}s_{a}s_{b}\right)$,
where $s^{2}\equiv\delta^{ab}s_{a}s_{b}>0$. Solving for $\delta_{ab}$,
we find $\delta_{ab}=\frac{1}{s^{2}}\eta_{ab}+\frac{2}{s^{2}}s_{a}s_{b}$.
Similar relations hold for the inverses, $\eta^{ab},\delta^{ab}$,
see Appendix 2. In addition, we may invert the basis change to write
the coordinate differentials,
\begin{eqnarray*}
\mathbf{d}w^{\beta} & = & k\beta h_{a}^{\;\beta}\left(\mathbf{e}^{a}-k\eta^{ab}\mathbf{f}_{b}\right)\\
\mathbf{d}s_{\alpha} & = & \frac{1}{2\beta}h_{\alpha}^{\; a}\left(\left(1-k\beta^{2}\right)\eta_{ab}\mathbf{e}^{b}+k\left(1+k\beta^{2}\right)\mathbf{f}_{a}\right)
\end{eqnarray*}
The known solution for the spin connection and Weyl form, eqs.(\ref{eq:original solution spin connection},\ref{original solution weyl vector})
immediately become
\begin{eqnarray}
\boldsymbol{\omega}_{\; b}^{a} & = & 2\Delta_{db}^{ac}s_{c}k\beta\left(\mathbf{e}^{d}-k\eta^{de}\mathbf{f}_{e}\right)\label{Solution full spin connection}\\
\boldsymbol{\omega} & = & -k\beta s_{a}\mathbf{e}^{a}+\beta\eta^{ab}s_{a}\mathbf{f}_{b}\label{Solution Weyl vector}
\end{eqnarray}
where we easily expand the projection $\Delta_{db}^{ac}$ in terms
of the new metric. Substituting this expansion to find $\boldsymbol{\tau}_{\; b}^{a}$,
results in
\begin{eqnarray*}
\boldsymbol{\tau}_{\; b}^{a} & = & \beta\left(2\Theta_{db}^{ac}s_{c}+2\eta^{ae}\eta_{bd}s_{e}+2\eta^{ae}s_{e}s_{b}s_{d}\right)\left(k\mathbf{e}^{d}-\eta^{dg}\mathbf{f}_{g}\right)-h_{b}^{\;\alpha}\mathbf{d}h_{\alpha}^{\; a}
\end{eqnarray*}
The antisymmetric part is then $\boldsymbol{\alpha}_{\; b}^{a}\equiv\Theta_{cb}^{ad}\boldsymbol{\tau}_{\; d}^{c}=-\Theta_{cb}^{ad}h_{d}^{\;\alpha}\mathbf{d}h_{\alpha}^{\; c}$
with the remaining terms cancelling identically. Furthermore, as described
above, $h_{\alpha}^{\; c}$ is a purely $s_{\alpha}$-dependent rotation
at each point. Therefore the remaining $h_{d}^{\;\alpha}\mathbf{d}h_{\alpha}^{\; c}$
term will lie totally in the subspace spanned by $\mathbf{d}s_{\alpha}$,
giving the parts of $\boldsymbol{\alpha}_{\; b}^{a}$ as 
\begin{eqnarray}
\boldsymbol{\sigma}_{\; b}^{a} & = & -\frac{1-k\beta^{2}}{2\beta}\Theta_{cb}^{ad}\left(h_{b}^{\;\alpha}\frac{\partial}{\partial s_{\beta}}h_{\alpha}^{\; a}\right)h_{\beta}^{\; c}\eta_{cd}\mathbf{e}^{d}\label{Sigma solution}\\
\boldsymbol{\gamma}_{\; b}^{a} & = & -\frac{k+\beta^{2}}{2\beta}\Theta_{cb}^{ad}\left(h_{b}^{\;\alpha}\frac{\partial}{\partial s_{\beta}}h_{\alpha}^{\; a}\right)h_{\beta}^{\; c}\mathbf{f}_{c}\label{Gamma solution}
\end{eqnarray}
Recall that the value of $k$ or $\beta$ in these expressions is
essentially a gauge choice and should be physically irrelavant. If
we choose $\beta^{2}=1$, we get either $\boldsymbol{\sigma}_{\; b}^{a}=0$
or $\boldsymbol{\gamma}_{\; b}^{a}=0$, depending on the sign of $k$.

Continuing, we are particularly interested in the symmetric pieces
of the connection since they constitute a new feature of the theory.
Applying the symmetric projection to $\boldsymbol{\tau}_{\; b}^{a}$,
we expand
\begin{eqnarray*}
\boldsymbol{\beta}_{\; b}^{a} & \equiv & \Xi_{cb}^{ad}\boldsymbol{\tau}_{\; d}^{c}
\end{eqnarray*}
Using $\Xi_{ab}^{cd}\left(h_{d}^{\;\mu}\mathbf{d}h_{\mu}^{\; a}\right)=\frac{1}{2}h_{\alpha}^{\; c}h_{b}^{\;\beta}k^{\alpha\mu}\mathbf{d}k_{\mu\beta}$
(see Appendix 3) to express the derivative term in terms of $v_{a}$,
we find the independent parts
\begin{eqnarray*}
\boldsymbol{\mu}_{\; b}^{a} & = & \left(-k\beta\delta_{b}^{a}s_{c}+\beta\gamma_{+}\left(\delta_{b}^{a}s_{c}+\delta_{c}^{a}s_{b}+\eta^{ad}\eta_{bc}s_{d}+2\eta^{ad}s_{b}s_{c}s_{d}\right)\right)\mathbf{e}^{c}\\
\boldsymbol{\rho}_{\; b}^{a} & = & \left(\beta\delta_{b}^{a}\eta^{cd}s_{d}+k\beta\gamma_{-}\left(\delta_{b}^{a}\eta^{cd}s_{d}+\delta_{b}^{c}\eta^{ad}s_{d}+\eta^{ac}s_{b}+2\eta^{ad}\eta^{ce}s_{b}s_{d}s_{e}\right)\right)\mathbf{f}_{c}
\end{eqnarray*}
where $\gamma_{\pm}\equiv\frac{1}{2\beta}\left(1\pm k\beta^{2}\right)$.
Written in this form, the tensor character of $\boldsymbol{\mu}_{\; b}^{a}$
and $\boldsymbol{\rho}_{\; b}^{a}$ is not evident, but since we have
chosen $\eta_{ab}$ orthonormal (referred to later as the orthonormal
gauge), $\phi=0$, and $\mathbf{v}=\boldsymbol{\omega}+\mathbf{d}\phi=\boldsymbol{\omega}$
we have $\mathbf{v}_{\left(e\right)}+\mathbf{u}_{\left(f\right)}=-k\beta s_{a}\mathbf{e}^{a}+\beta\eta^{ab}s_{a}\mathbf{f}_{b}$
so that we may equally well write

\begin{eqnarray}
\boldsymbol{\mu}_{\; b}^{a} & = & \left(\delta_{b}^{a}v_{c}-k\gamma_{+}\left(\delta_{b}^{a}v_{c}+\delta_{c}^{a}v_{b}+\eta^{ad}\eta_{bc}v_{d}+\frac{2}{\beta^{2}}\eta^{ad}v_{b}v_{c}v_{d}\right)\right)\mathbf{e}^{c}\label{eq:NewSymmSpinConnectionMu}\\
\boldsymbol{\rho}_{\; b}^{a} & = & \left(\delta_{b}^{a}u^{c}+k\gamma_{-}\left(\delta_{b}^{a}u^{c}+\delta_{b}^{c}u^{a}+\eta^{ac}\eta_{bd}u^{d}+\frac{2}{\beta^{2}}\eta_{bd}u^{a}u^{c}u^{d}\right)\right)\mathbf{f}_{c}\label{eq:NewSymmSpinConnectionRho}
\end{eqnarray}
which are manifestly tensorial.

The involution conditions, eqs.(\ref{eq:InvolutionE}-\ref{eq:InvolutionF}),
are easily seen to be satisfied identically by eqs.(\ref{eq:NewSymmSpinConnectionMu},
\ref{eq:NewSymmSpinConnectionRho}). Therefore, the $\mathbf{f}_{a}=0$
and $\mathbf{e}^{a}=0$ subspaces are Lagrangian submanifolds spanned
respectively by $\mathbf{e}^{a}$ and $\mathbf{f}_{a}$. There exist
coordinates such that these basis forms may be written
\begin{eqnarray}
\mathbf{e}^{a} & = & e_{\mu}^{\quad a}\mathbf{d}x^{\mu}\label{Solder form in adapted coords}\\
\mathbf{f}_{a} & = & f_{a}^{\quad\mu}\mathbf{d}y_{\mu}\label{Co-solder form in adapted coordinates}
\end{eqnarray}
To find such submanifold coordinates, the useful thing to note is
that $\mathbf{d}\left(\frac{s_{\alpha}}{s^{2}}\right)=\delta_{\alpha\nu}k^{\mu\nu}\mathbf{d}s_{\mu}$
so that the basis may be written as
\begin{eqnarray*}
\mathbf{e}^{a} & = & h_{\alpha}^{\; a}\mathbf{d}\left(k\gamma_{+}w^{\alpha}+\beta\delta^{\alpha\beta}\left(\frac{s_{\alpha}}{s^{2}}\right)\right)\equiv h_{\alpha}^{\; a}\mathbf{d}x^{\alpha}\\
\mathbf{f}_{a} & = & \left(h_{a}^{\;\alpha}k_{\alpha\beta}\delta^{\beta\mu}\right)\mathbf{d}\left(k\beta\left(\frac{s_{\mu}}{s^{2}}\right)-\gamma_{-}\delta_{\mu\nu}w^{\nu}\right)\equiv f_{a}^{\;\mu}\mathbf{d}y_{\mu}
\end{eqnarray*}
with $x^{\alpha}=k\gamma_{+}w^{\alpha}+\beta\delta^{\alpha\beta}\left(\frac{s_{\alpha}}{s^{2}}\right)$
and $y_{\mu}=k\beta\left(\frac{s_{\mu}}{s^{2}}\right)-\gamma_{-}\delta_{\mu\nu}w^{\nu}$.
This confirms the involution.

\section{Curvature of the submanifolds}

The nature of the configuration or momentum submanifold may be determined
by restricting the structure equations by $\mathbf{f}_{a}=0$ or $\mathbf{e}^{a}=0$,
respectively. To aid in the interpretation of the resulting submanifold
structure equations, we define the curvature of the antisymmetric
connection $\boldsymbol{\alpha}_{\: b}^{a}$

\begin{eqnarray}
\mathbf{R}_{\; b}^{a} & \equiv & \mathbf{d}\boldsymbol{\alpha}_{\; b}^{a}-\boldsymbol{\alpha}_{\; b}^{c}\wedge\boldsymbol{\mathbf{\alpha}}_{\; c}^{a}\label{eq:RiemannCurvDef-1}\\
 & = & \frac{1}{2}R_{\; bcd}^{a}\mathbf{e}^{c}\wedge\mathbf{e}^{d}+R_{\; b\; d}^{a\; c}\mathbf{f}_{c}\wedge\mathbf{e}^{d}+\frac{1}{2}R_{\; b}^{a\; cd}\mathbf{f}_{c}\wedge\mathbf{f}_{d}
\end{eqnarray}
While all components of the overall Cartan curvature, $\boldsymbol{\Omega}^{A}=\left(\boldsymbol{\Omega}_{b}^{a},\mathbf{T}^{a},\mathbf{S}_{a},\boldsymbol{\Omega}\right)$
are zero on $\mathcal{M}_{0}^{\left(2n\right)}$, the curvature, $\mathbf{R}_{\; b}^{a}$,
and in particular the curvatures $\frac{1}{2}R_{\; bcd}^{a}\mathbf{e}^{c}\wedge\mathbf{e}^{d}$
and $\frac{1}{2}R_{\; b}^{a\; cd}\mathbf{f}_{c}\wedge\mathbf{f}_{d}$
on the submanifolds, may or may not be. Here we examine this question,
using the structure equations to find the \emph{Riemannian} curvature
of the connections, $\boldsymbol{\sigma}_{b}^{a}$ and $\boldsymbol{\gamma}_{b}^{a},$
of the Lorentzian submanifolds.

\subsection{Momentum space curvature\label{sub:SpencWheelerMetric}}

To see that the Lagrangian submanifold equations describe a Riemannian
geometry, we set $\mathbf{e}^{a}=0$ in the structure equations, eqs.(\ref{eq:NewStrucSpinConn}-\ref{eq:NewStrucWeyl})
and choose the $\phi=0$ (orthonormal) gauge (or see Appendix 1, eqs.(\ref{eq:MomCurv}-\ref{eq:MomDil},
with the Cartan curvatures set to zero). Then, taking the $\Theta_{db}^{ac}$
projection, we have
\begin{eqnarray}
0 & = & \frac{1}{2}R_{\; b}^{a\; cd}\mathbf{f}_{c}\wedge\mathbf{f}_{d}-\boldsymbol{\rho}_{\; b}^{c}\wedge\boldsymbol{\mathbf{\rho}}_{\; c}^{a}+\Theta_{db}^{ac}\eta^{ac}\Delta_{cf}^{eb}\mathbf{f}_{b}\wedge\mathbf{f}_{a}\label{eq:MomentumRiemannCurv}\\
0 & = & \mathbf{d}_{\left(y\right)}\mathbf{f}_{b}-\boldsymbol{\gamma}_{\; b}^{a}\wedge\mathbf{f}_{a}\nonumber 
\end{eqnarray}
These are the structure equations of a Riemannian geometry with additional
\emph{geometric} terms, $-\boldsymbol{\rho}_{\; b}^{c}\wedge\boldsymbol{\mathbf{\rho}}_{\; c}^{a}+\Theta_{db}^{ac}\eta^{ac}\Delta_{cf}^{eb}\mathbf{f}_{b}\wedge\mathbf{f}_{a}$,
reflecting the difference between Lorentz curvature and conformal
curvature. The symmetric projection is
\begin{eqnarray*}
\mathbf{D}^{\left(y\right)}\boldsymbol{\rho}_{\; b}^{a} & = & -k\Xi_{db}^{ac}\Delta_{ec}^{df}\eta^{eg}\mathbf{f}_{f}\wedge\mathbf{f}_{g}\\
\mathbf{d}_{\left(y\right)}\mathbf{u}_{\left(f\right)} & = & 0
\end{eqnarray*}
where $\mathbf{u}_{\left(f\right)},\boldsymbol{\gamma}_{\; b}^{a}$
and $\boldsymbol{\rho}_{\; b}^{a}$ are given by eqs.(\ref{Solution Weyl vector},\ref{Gamma solution},\ref{eq:NewSymmSpinConnectionRho}),
respectively. Rather than computing $R_{\; b}^{a\; cd}$ directly
from $\boldsymbol{\gamma}_{\; b}^{a}$, which requires a complicated
expression for the local rotation, $h_{\alpha}^{\; a}$, we find it
using the rest of eq. (\ref{eq:MomentumRiemannCurv}). 

Letting $\beta=e^{\lambda}$ so that
\begin{eqnarray*}
k+\gamma_{-}^{2} & = & \left\{ \begin{array}{cc}
\cosh^{2}\lambda & k=1\\
\sinh^{2}\lambda & k=-1
\end{array}\right.
\end{eqnarray*}
the curvature is
\begin{eqnarray*}
\frac{1}{2}R_{\; b}^{a\; cd}\mathbf{f}_{c}\mathbf{f}_{d} & = & \left\{ \begin{array}{cc}
\cosh^{2}\lambda\Theta_{cb}^{ag}\left(\eta^{cf}+2\eta^{cd}\eta^{fe}s_{d}s_{e}\right)\mathbf{f}_{f}\wedge\mathbf{f}_{g} & k=1\\
\sinh^{2}\lambda\Theta_{cb}^{ag}\left(\eta^{cf}+2\eta^{cd}\eta^{fe}s_{d}s_{e}\right)\mathbf{f}_{f}\wedge\mathbf{f}_{g} & k=-1
\end{array}\right.
\end{eqnarray*}
Now consider the symmetric equations. Notice that the Weyl vector
has totally decoupled, with its equation showing that $\mathbf{u}_{\left(f\right)}$
is closed, a result which also follows from its definition. For the
symmetric projection, we find $\Xi_{db}^{ac}\eta^{ac}\Delta_{cf}^{eb}\mathbf{f}_{b}\mathbf{f}_{a}\equiv0$.
Then, contraction of $D^{a}\rho_{\; b}^{a\;\; c}$ with $\eta_{ad}\eta_{ce}u^{a}u^{e}$
, together with $\mathbf{d}_{\left(y\right)}\mathbf{u}_{\left(f\right)}=0$
shows that $u^{a}$ is covariantly constant, $D_{\left(y\right)}^{a}u^{b}=0$.

If we choose $k=-1$ and $\lambda=0$, the Riemann curvature of the
momentum space vanishes. This is a stronger result than in \cite{Spencer:2008p167},
since there only the Weyl curvature could be set to zero consistently.
In this case, the Lagrangian submanifold becomes a vector space and
there is a natural interpretation as the co-tangent space of the configuration
space. However, the orthonormal metric in this case, $\left\langle \mathbf{f}_{a},\mathbf{f}_{b}\right\rangle =\eta_{ab}$,
has the opposite sign from the metric of the configuration space,
$\left\langle \mathbf{e}^{a},\mathbf{e}^{b}\right\rangle =-\eta^{ab}$.
This reversal of sign of the metric together with the the units, suggests
that the physical (momentum) tangent space coordinates are related
to the geometrical ones by $p_{\alpha}\sim i\hbar y_{\alpha}$. This
has been suggested previously \cite{Wheeler:1997tg} and explored
in the context of quantization \cite{Anderson:2004p612}.

Leaving $\beta$ and $k$ unspecified, we see that in general momentum
space has non-vanishing Riemannian curvature of the connection $\boldsymbol{\gamma}_{\; b}^{a}$,
a situation suggested long ago for quantum gravity \cite{Born:1938p217,Freidel:2006p21}.
We consider this further in Section 7.3. Whatever the values of $\beta$
and $k$, the momentum space is conformally flat. We see this from
the decomposition of Riemannian curvature into the Weyl curvature,
$\mathbf{C}_{\; b}^{a}$, and Schouten tensor, $\boldsymbol{\mathcal{R}}_{a}$,
given by
\begin{eqnarray*}
\mathbf{R}_{\; b}^{a} & = & \mathbf{C}_{\; b}^{a}-2\Theta_{db}^{ae}\boldsymbol{\mathcal{R}}_{e}\mathbf{e}^{d}
\end{eqnarray*}
The Schouten tensor,$\boldsymbol{\mathcal{R}}_{a}\equiv\frac{1}{n-2}\left(R_{ab}-\frac{1}{2\left(n-1\right)}R\eta_{ab}\right)\mathbf{e}^{b}$
is algebraically equivalent to the Ricci tensor, $R_{ab}$. It is
easy to prove that when the curvature $2$-form can be expressed as
a projection in the form $\mathbf{R}_{\; b}^{a}=-2\Theta_{db}^{ae}\mathbf{X}_{e}\mathbf{e}^{d}$,
then $\mathbf{X}_{a}$ is the Schouten tensor, and the Weyl curvature
vanishes. Vanishing Weyl curvature implies conformal flatness.

\subsection{Spacetime curvature and geometric curvature}

The curvature on the configuration space takes the same basic form.
Still in the orthonormal gauge, and separating the symmetric and antisymmetric
parts as before, we again find a Riemannian geometry with additional
geometric terms,
\begin{eqnarray}
0 & = & \mathbf{R}_{\; b}^{a}\left(\boldsymbol{\sigma}\right)-\boldsymbol{\mu}_{\; b}^{c}\boldsymbol{\mu}_{\; c}^{a}-\Theta_{db}^{ac}\Delta_{fc}^{de}\eta_{eg}\mathbf{e}^{g}\mathbf{e}^{f}\label{Configuration structure eq curvature}\\
0 & = & \mathbf{d}_{\left(x\right)}\mathbf{e}^{a}-\mathbf{e}^{b}\boldsymbol{\sigma}_{\; b}^{a}\label{Configuration structure eq solder}
\end{eqnarray}
together with
\begin{eqnarray*}
0 & = & \mathbf{D}_{\left(x\right)}\boldsymbol{\mu}_{\; b}^{a}-\Xi_{db}^{ac}\Delta_{fc}^{de}\eta_{eg}\mathbf{e}^{g}\mathbf{e}^{f}\\
0 & = & \mathbf{d}_{\left(x\right)}\mathbf{v}
\end{eqnarray*}

Looking first at all the $\Theta_{cb}^{ad}$-antisymmetric terms and
substituting in (\ref{eq:NewSymmSpinConnectionMu}) for $\boldsymbol{\mu}_{\; b}^{a}$,
we find that the Riemannian curvature is 
\begin{eqnarray*}
\mathbf{R}_{\; b}^{a} & = & \left(\gamma_{+}^{2}-k\right)\Theta_{db}^{ac}\left(\eta_{ce}+2s_{c}s_{e}\right)\mathbf{e}^{d}\mathbf{e}^{e}
\end{eqnarray*}
so the Weyl curvature vanishes and the Schouten tensor is
\begin{eqnarray}
\boldsymbol{\mathcal{R}}_{a} & = & \frac{1}{2}\left(\gamma_{+}^{2}-k\right)\left(\eta_{ab}+2s_{a}s_{b}\right)\mathbf{e}^{b}\label{Configuration Schouten}
\end{eqnarray}
The vanishing Weyl curvature tensor shows that the spacetime is conformally
flat. This result is discussed in detail below.

The equation, $\mathbf{d}_{\left(x\right)}\mathbf{v}=0$ shows that
$\mathbf{v}$ is hypersurface orthogonal. Expanding the remaining
equation with $\mathbf{d}_{\left(x\right)}\mathbf{v}=0,\,\mathbf{D}_{\left(x\right)}\eta_{ab}=0$
and $\mathbf{D}_{\left(x\right)}\mathbf{e}^{a}=0$, contractions involving
$\eta_{ab}$ and $v_{a}$ quickly show that 
\[
D_{a}^{\left(x\right)}v_{b}=0
\]
This, combined with $\mathbf{D}^{\left(y\right)}u^{a}=0$ and $u^{a}=-k\eta^{ab}v_{b}$
shows that the full covariant derivative vanishes, $D_{a}v_{b}=0$.
The scale vector is therefore a covariantly constant, hypersurface
orthogonal, unit timelike Killing vector of the spacetime submanifold.

\subsection{Curvature invariant}

Substituting $\beta=e^{\lambda}$ as before, the components of the
momentum and configuration curvatures become
\[
\eta_{df}\eta_{eg}R_{\; b}^{a\; fg}=\left\{ \begin{array}{cc}
\cosh^{2}\lambda\left(\Theta_{db}^{ac}\delta_{e}^{f}-\Theta_{eb}^{ac}\delta_{d}^{f}\right)\left(\eta_{fc}+2s_{f}s_{c}\right) & k=1\\
\sinh^{2}\lambda\left(\Theta_{db}^{ac}\delta_{e}^{f}-\Theta_{eb}^{ac}\delta_{d}^{f}\right)\left(\eta_{fc}+2s_{f}s_{c}\right) & k=-1
\end{array}\right.
\]
and 
\begin{eqnarray*}
R_{\; bde}^{a} & = & \left\{ \begin{array}{cc}
\sinh^{2}\lambda\left(\Theta_{db}^{ac}\delta_{e}^{f}-\Theta_{eb}^{ac}\delta_{d}^{f}\right)\left(\eta_{fc}+2s_{f}s_{c}\right) & k=1\\
\cosh^{2}\lambda\left(\Theta_{db}^{ac}\delta_{e}^{f}-\Theta_{eb}^{ac}\delta_{d}^{f}\right)\left(\eta_{fc}+2s_{f}s_{c}\right) & k=-1
\end{array}\right.
\end{eqnarray*}
Subtracting these 
\begin{eqnarray*}
\eta_{df}\eta_{eg}R_{\; b}^{a\; fg}-R_{\; bde}^{a} & = & k\left(\Theta_{db}^{ac}\delta_{e}^{f}-\Theta_{eb}^{ac}\delta_{d}^{f}\right)\left(\eta_{fc}+2s_{f}s_{c}\right)
\end{eqnarray*}
so that the difference of the configuration and momentum curvatures
is independent of the linear combination of basis forms used. This
coupling between the momentum and configuration space curvatures adds
a sort of complementarity that goes beyond the suggestion by Born
\cite{Born:1938p217,Freidel:2006p21} that momentum space might also
be curved. As we continuously vary $\beta^{2}$, the curvature moves
between momentum and configuration space but this difference remains
unchanged. We may even make one or the other Lagrangian submanifold
flat.

For the Einstein tensors,
\begin{eqnarray*}
\eta_{ac}\eta_{bd}G_{\left(y\right)}^{cd}-G_{ab}^{\left(x\right)} & = & \frac{1}{2}k\left(\left(n-3\right)\eta_{ab}+\left(n-2\right)s_{a}s_{b}\right)
\end{eqnarray*}

\subsection{Candidate dark matter}

There is a surprising consequence of the tensor $\boldsymbol{\mu}_{\; b}^{a}$
in the Lorentz structure equation. The structure equations for the
configuration Lagrangian submanifold above describe an ordinary curved
Lorentzian spacetime with certain extra terms from the conformal geometry
that exist even in the \emph{absence of matter}. We gain some insight
into the nature of these additional terms from the metric and Einstein
tensor. In coordinates, the metric takes the form
\[
h_{\alpha\beta}=s^{2}\left(\delta_{\alpha\beta}-\frac{2}{s^{2}}s_{\alpha}s_{\beta}\right)
\]
which is straightforwardly boosted to $s^{2}\eta_{\alpha\beta}^{0}$
at a point. Since the spacetime is conformally flat, gradients of
the conformal factor must be in the time direction, $s_{\alpha}$,
so we may rescale the time, $dt'=\sqrt{s^{2}}dt$ to put the line
element in the form
\[
ds^{2}=-dt'^{2}+s^{2}\left(t'\right)\left(dx^{2}+dy^{2}+dz^{2}\right)
\]
That is, the \emph{vacuum solution} is a spatially flat FRW cosmology.
Putting the results in terms of the Einstein tensor and a coordinate
basis, we expect an equation of the form $\tilde{G}_{\alpha\beta}=\kappa T_{\alpha\beta}^{matter}$
where the Cartan Einstein tensor is modified to
\begin{eqnarray}
\tilde{G}_{\alpha\beta} & \equiv & G_{\alpha\beta}-3\left(n-2\right)s^{2}s_{\alpha}s_{\beta}+\frac{3}{2}\left(n-2\right)\left(n-3\right)s^{2}h_{\alpha\beta}\label{eq:EinsteinTensor-1}
\end{eqnarray}
where $G_{\alpha\beta}$ is the familiar Einstein tensor. The new
geometric terms may be thought of as a combination of a \emph{cosmological
constant} and a \emph{cosmological perfect fluid}. With this interpretation,
we may write the new cosmological terms as
\begin{eqnarray*}
\kappa T_{\alpha\beta}^{cosm} & = & \left(\rho_{0}+p_{0}\right)v_{\alpha}v_{\beta}+p_{0}h_{\alpha\beta}-\Lambda h_{\alpha\beta}
\end{eqnarray*}
where $\kappa T_{ab}^{cosm}\equiv3\left(n-2\right)s^{2}v_{\alpha}v_{\beta}-\frac{3}{2}\left(n-2\right)\left(n-3\right)s^{2}h_{\alpha\beta}$.
In $n=4$-dimensions, $\frac{1}{2}\left(\rho_{0}+p_{0}\right)=\Lambda-p_{0}$,
with the equation of state and the overall scale undetermined. If
we assume an equation of state $p_{0}=w\rho_{0}$, this becomes
\[
\frac{1}{2}\left(1+3w\right)\rho_{0}=\Lambda
\]
This relation alone does not account for the values suggested by the
current Planck data: about $0.68$ for the cosmological constant,
$0.268$ for the density of dark matter, and vanishing pressure, $w=0$.
However, these values are based on standard cosmology, while we have
not yet included matter terms in eq.(\ref{eq:EinsteinTensor-1}).
Moreover, the proportions of the three geometric terms in eq.(\ref{eq:EinsteinTensor-1})
may change when curvature is included. Such a change is suggested
by the form of known solutions in the original basis, where $h_{\alpha\beta}$
is augmented by a Schouten term. If this modification also occurs
in the adapted basis, the ratios above will be modified. We are currently
examining such solutions.

\section{Discussion}

Using the quotient method of gauging, we constructed the class of
biconformal geometries . The construction starts with the conformal
group of an $SO\left(p,q\right)$-symmetric pseudo-metric space. The
quotient by $\mathcal{W}\left(p,q\right)\equiv SO\left(p,q\right)\times dilatations$
gives the homogeneous manifold, $\mathcal{M}_{0}^{2n}$. We show that
this manifold is metric and symplectic (as well as Kähler with a different
metric). Generalizing the manifold and connection while maintaining
the local $\mathcal{W}$ invariance, we display the resulting biconformal
spaces, $\mathcal{M}^{2n}$ \cite{Ivanov:1982p1172,Ivanov:1982p1201,Wheeler:1997pc}.

This class of locally symmetric manifolds becomes a model for gravity
when we recall the most general curvature-linear action \cite{Wehner:1999p1653}.

It is shown in \cite{Spencer:2008p167} that $\mathcal{M}_{0}^{\left(2n\right)}\left(p,q\right)$
in any dimension $n=p+q$ will have\emph{ Lagrangian submanifolds
that are orthogonal with respect to the 2n-dim biconformal (Killing)
metric and have non-degenerate $n$-dim metric restrictions on those
submanifolds} only if the original space is Euclidean or signature
zero $\left(p\in\left\{ 0,\frac{n}{2},n\right\} \right)$, and then
the signature of the submanifolds is severely limited $\left(p\rightarrow p\pm1\right)$.
This leads in the two Euclidean cases to Lorenztian configuration
space, and hence the origin of time \cite{Spencer:2008p167}. For
the case of flat, $8$-dim biconformal space the Lagrangian submanifolds
are necessarily Lorentzian.

Our investigation explores properties of the homogeneous manifold,
$\mathcal{M}_{0}^{2n}\left(n,0\right)$. Starting with Euclidean symmetry,
$SO\left(n\right)$, we clarify the emergence of Lorentzian signature
Lagrangian submanifolds. We extend the results of \cite{Spencer:2008p167},
eliminating all but the group-theoretic assumptions. By writing the
structure equations in an adapted basis, we reveal new features of
these geometries. We summarize our new findings below.

\textbf{A new connection}

There is a natural $SO\left(n\right)$ Cartan connection on $\mathcal{M}_{0}^{2n}$.
Rewriting the biconformal structure equations in an orthogonal, canonically
conjugate, conformally orthonormal basis automatically introduces
a Lorentzian connection and decouples the Weyl vector from the submanifolds.
This structure emerges directly from the transformation of the structure
equations, as detailed in points $1$ through $4$ in $\S$\ref{sub:OrthonormalProperties}. 

Specifically, we showed that all occurences of the $SO\left(4\right)$
spin connection $\boldsymbol{\omega}_{\;\beta}^{\alpha}$ may be written
in terms of the new connection, $\boldsymbol{\tau}_{\; b}^{a}\equiv h_{\alpha}^{\; a}\boldsymbol{\omega}_{\;\beta}^{\alpha}h_{b}^{\;\beta}-h_{b}^{\;\alpha}\mathbf{d}h_{\alpha}^{\; a}$,
which has both symmetric and antisymmetric parts. These symmetric
and antisymmetric parts separate \emph{automatically} in the structure
equations, with only the Lorentz part of the connection, $\boldsymbol{\alpha}_{\; b}^{a}=\Theta_{db}^{ac}\boldsymbol{\tau}_{\; c}^{d}$
describing the evolution of the configuration submanifold solder form.
The spacetime and momentum space connections are metric compatible,
up to a conformal factor.

The Weyl vector terms drop out of the submanifold basis equations.
The submanifold equations remain scale invariant because of the residual
metric derivative, $\frac{1}{2}\mathbf{d}\eta^{ac}\eta_{cb}=\delta_{b}^{a}\mathbf{d}\phi$.
When the metric is rescaled, this term changes with the negative of
the inhomogeneous term acquired by the Weyl vector.

\textbf{Two new tensors}

It is especially striking how the Weyl vector and the symmetric piece
of the connection are pushed from the basis submanifolds into the
mixed basis equations. These extra degrees of freedom are embodied
in two new \emph{Lorentz tensors}.

The factor $\delta_{b}^{a}\mathbf{d}\phi$ which replaces the Weyl
vector in the basis equations allows us to form a \emph{scale-invariant}
$1$-form,
\[
\mathbf{v}=\boldsymbol{\omega}+\mathbf{d}\phi
\]
It is ultimately this vector which determines the time direction.

We showed that the symmetric part of the spin connection, $\boldsymbol{\beta}_{\; b}^{a}$,
despite being a piece of the connection, transforms as a tensor. The
solution of the structure equations shows that the two tensors, $\mathbf{v}$
and $\boldsymbol{\beta}_{\; b}^{a}$ are related, with $\boldsymbol{\beta}_{\; b}^{a}$
constructed cubically, purely from $\mathbf{v}$ and the metric. Although
the presence of $\boldsymbol{\beta}_{\; b}^{a}$ changes the form
of the momentum space curvature, we find the same signature changing
metric as found in \cite{Spencer:2008p167}. Rather than imposing
vanishing momentum space curvature as in \cite{Spencer:2008p167},
we make use of a complete solution of the Maurer-Cartan equations
to derive the metric. The integrability of the Lagrangian submanifolds,
the Lorentzian metric and connection, and the possibility of a flat
momentum space are all now seen as direct consequences of the structure
equations, without assumptions.

\textbf{Riemannian spacetime and momentum space}

The configuration and momentum submanifolds have vanishing dilatational
curvature, making them gauge equivalent to Riemannian geometries.
Together with the signature change from the original Euclidean space
to these Lorentzian manifolds, we arrive at a suitable arena for general
relativity in which time is constructed covariantly from a scale-invariant
Killing field. This field is provided automatically from the group
structure.

\textbf{Effective cosmological fluid and cosmological constant}

Though we work in the homogeneous space, $\mathcal{M}_{0}^{2n}$,
so that there are no Cartan curvatures, there is a net Riemannian
curvature remaining on the spacetime submanifold. We show this to
describe a conformally flat spacetime with the deviation from flatness
provided by additional geometric terms of the form
\[
\tilde{G}_{\alpha\beta}\equiv G_{\alpha\beta}-\rho_{0}v_{\alpha}v_{\beta}+\Lambda h_{\alpha\beta}=0
\]
that is, a background dust and a cosmological constant. The values
$\rho_{0}=3\left(n-2\right)s^{2}$ and $\Lambda=\frac{3}{2}\left(n-2\right)\left(n-3\right)s^{2}$
give, in the absence of physical sources, the relation $\left(2+3w\right)\rho_{0}=\Lambda$
for an equation of state $p_{0}=w\rho_{0}$. An examination of more
realistic cosmological models involving matter fields and curved biconformal
spaces, $\mathcal{M}^{2n}$, is underway. 

\appendix

\section*{Appendix 1: Subparts of the structure equations}

\setcounter{section}{1}

Here we write the structure equations, including Cartan curvature.
We expand the configuration, mixed and momentum terms separately.
Note that the $\mathbf{f}_{a}\mathbf{f}_{b}$ part of the $\mathbf{d}\mathbf{e}^{a}$
equation and the $\mathbf{e}^{a}\mathbf{e}^{b}$ part of the $\mathbf{d}\mathbf{f}_{a}$
equation are set to zero. These are the involution conditions, which
guarantee that the configuration and momentum subspaces are integrable
submanifolds by the Frobenius theorem.

In the conformal-orthonormal basis, we have $g^{ab}\mathbf{d}g_{bc}=e^{-2\phi}\eta^{ab}\mathbf{d}\left(e^{2\phi}\eta_{bc}\right)=2\delta_{c}^{a}\mathbf{d}\phi$.
The structure equations in the conformal-orthonormal basis are
\begin{eqnarray*}
\mathbf{d}\boldsymbol{\tau}_{\; b}^{a} & = & \boldsymbol{\tau}_{\; b}^{c}\boldsymbol{\mathbf{\tau}}_{\; c}^{a}+\Delta_{db}^{ae}\eta_{ec}\mathbf{e}^{c}\mathbf{e}^{d}-\Delta_{eb}^{ac}\eta^{ed}\mathbf{f}_{c}\mathbf{f}_{d}+2\Delta_{fb}^{ae}\Xi_{de}^{fc}\mathbf{f}_{c}\mathbf{e}^{d}+\boldsymbol{\Omega}_{\; b}^{a}\\
\mathbf{d}\mathbf{e}^{a} & = & \mathbf{e}^{c}\boldsymbol{\alpha}_{\; c}^{a}+\frac{1}{2}\eta_{cb}\mathbf{d}\eta^{ac}\mathbf{e}^{b}+\frac{1}{2}\mathbf{D}\eta^{ab}\mathbf{f}_{b}+\mathbf{T}^{a}\\
\mathbf{d}\mathbf{f}_{a} & = & \boldsymbol{\alpha}_{\; a}^{b}\mathbf{f}_{b}+\frac{1}{2}\eta^{bc}\mathbf{d}\eta_{ab}\mathbf{f}_{c}-\frac{1}{2}\mathbf{D}\eta_{ab}\mathbf{e}^{b}+\mathbf{S}_{a}\\
\mathbf{d}\boldsymbol{\omega} & = & \mathbf{e}^{a}\mathbf{f}_{a}+\boldsymbol{\Omega}
\end{eqnarray*}
Then defining
\begin{eqnarray*}
\mathbf{D}^{\left(x\right)}\boldsymbol{\mu}_{\; b}^{a} & \equiv & \mathbf{d}^{\left(x\right)}\boldsymbol{\mu}_{\; b}^{a}-\boldsymbol{\mu}_{\; b}^{c}\boldsymbol{\sigma}_{\; c}^{a}-\boldsymbol{\sigma}_{\; b}^{c}\boldsymbol{\mu}_{\; c}^{a}\\
\mathbf{D}^{\left(x\right)}\boldsymbol{\rho}_{\; b}^{a} & \equiv & \mathbf{d}^{\left(x\right)}\boldsymbol{\rho}_{\; b}^{a}-\boldsymbol{\rho}_{\; b}^{c}\boldsymbol{\sigma}_{\; c}^{a}-\boldsymbol{\sigma}_{\; b}^{c}\boldsymbol{\rho}_{\; c}^{a}\\
\mathbf{D}^{\left(y\right)}\boldsymbol{\mu}_{\; b}^{a} & \equiv & \mathbf{d}^{\left(y\right)}\boldsymbol{\mu}_{\; b}^{a}-\boldsymbol{\mu}_{\; b}^{c}\boldsymbol{\gamma}_{\; c}^{a}-\boldsymbol{\gamma}_{\; b}^{c}\boldsymbol{\mu}_{\; c}^{a}\\
\mathbf{D}^{\left(y\right)}\boldsymbol{\rho}_{\; b}^{a} & \equiv & \mathbf{d}^{\left(y\right)}\boldsymbol{\rho}_{\; b}^{a}-\boldsymbol{\rho}_{\; b}^{c}\boldsymbol{\gamma}_{\; c}^{a}-\boldsymbol{\gamma}_{\; b}^{c}\boldsymbol{\rho}_{\; c}^{a}
\end{eqnarray*}
the separation of the structure equations into independent parts gives:

\subsubsection*{Configuration space:}

\numparts

\begin{eqnarray}
\frac{1}{2}\Omega_{\; bcd}^{a}\mathbf{e}^{c}\mathbf{e}^{d} & = & \mathbf{d}^{\left(x\right)}\boldsymbol{\sigma}_{\; b}^{a}-\boldsymbol{\sigma}_{\; b}^{c}\boldsymbol{\sigma}_{\; c}^{a}+\mathbf{D}^{\left(x\right)}\boldsymbol{\mu}_{\; b}^{a}-\boldsymbol{\mu}_{\; b}^{c}\boldsymbol{\mu}_{\; c}^{a}-k\Delta_{eb}^{ac}\eta_{cd}\mathbf{e}^{d}\mathbf{e}^{e}\label{eq:ConfiqCurv}\\
\frac{1}{2}T_{\; bc}^{a}\mathbf{e}^{b}\mathbf{e}^{c} & = & \mathbf{d}_{\left(x\right)}\mathbf{e}^{a}-\mathbf{e}^{b}\boldsymbol{\sigma}_{\; b}^{a}+\frac{1}{2}\eta^{ac}\mathbf{d}^{\left(x\right)}\eta_{cb}\mathbf{e}^{b}\label{eq:ConfigBasis}\\
\frac{1}{2}S_{abc}\mathbf{e}^{b}\mathbf{e}^{c} & = & k\eta_{ab}\mathbf{e}^{c}\left(\boldsymbol{\mu}_{\; c}^{b}-\delta_{c}^{b}W_{d}\mathbf{e}^{d}+\frac{1}{2}\eta_{ce}\mathbf{d}^{\left(x\right)}\eta^{be}\right)\label{eq:ConfigInv}\\
\frac{1}{2}\Omega_{ab}\mathbf{e}^{a}\mathbf{e}^{b} & = & \mathbf{d}_{\left(x\right)}\left(W_{a}\mathbf{e}^{a}\right)\label{eq:ConfigDil}
\end{eqnarray}
\endnumparts

\numparts

\subsubsection*{Cross-term:}

\begin{eqnarray}
\Omega_{\; b\; d}^{a\;\; c}\mathbf{f}_{c}\mathbf{e}^{d} & = & \mathbf{d}^{\left(y\right)}\boldsymbol{\sigma}_{\; b}^{a}+\mathbf{d}^{\left(x\right)}\boldsymbol{\gamma}_{\; b}^{a}-\boldsymbol{\gamma}_{\; b}^{c}\boldsymbol{\sigma}_{\; c}^{a}-\boldsymbol{\sigma}_{\; b}^{c}\boldsymbol{\gamma}_{\; c}^{a}\label{eq:MixedCurv}\\
 &  & +\mathbf{D}^{\left(x\right)}\boldsymbol{\rho}_{\; b}^{a}+\mathbf{D}^{\left(y\right)}\boldsymbol{\mu}_{\; b}^{a}-\boldsymbol{\rho}_{\; b}^{c}\boldsymbol{\mu}_{\; c}^{a}-\boldsymbol{\mu}_{\; b}^{c}\boldsymbol{\rho}_{\; c}^{a} \nonumber \\
 &  & -2\Delta_{db}^{ac}\Xi_{ce}^{fd}\mathbf{f}_{f}\mathbf{e}^{e}\nonumber \\
T_{\;\; c}^{ab}\mathbf{f}_{b}\mathbf{e}^{c} & = & \mathbf{d}^{\left(y\right)}\mathbf{e}^{a}-\mathbf{e}^{b}\boldsymbol{\gamma}_{\; b}^{a}+\frac{1}{2}\eta^{ac}\mathbf{d}^{\left(y\right)}\eta_{cb}\mathbf{e}^{b}\label{eq:MixedSolder}\\
 &  & -k\eta^{ac}\left(\boldsymbol{\mu}_{\; c}^{b}\mathbf{f}_{b}+W_{d}\mathbf{f}_{c}\mathbf{e}^{d}-\frac{1}{2}\eta^{bd}\mathbf{d}^{\left(x\right)}\eta_{cd}\mathbf{f}_{b}\right) \nonumber \\
S_{a\;\; c}^{\; b}\mathbf{f}_{b}\mathbf{e}^{c} & = & \mathbf{d}^{\left(x\right)}\mathbf{f}_{a}-\boldsymbol{\sigma}_{\; a}^{b}\mathbf{f}_{b}-\frac{1}{2}\eta^{cb}\mathbf{d}^{\left(x\right)}\eta_{ac}\mathbf{f}_{b}\label{MixedCosold}\\
 &  & +k\eta_{ab}\left(\mathbf{e}^{c}\boldsymbol{\rho}_{\; c}^{b}+W^{c}\mathbf{f}_{c}\mathbf{e}^{b}+\frac{1}{2}\eta^{bc}\mathbf{d}^{\left(y\right)}\eta_{cd}\mathbf{e}^{d}\right) \nonumber \\
\Omega_{\; b}^{a}\mathbf{f}_{a}\mathbf{e}^{b} & = & \mathbf{d}_{\left(y\right)}\left(W_{a}\mathbf{e}^{a}\right)+\mathbf{d}_{\left(x\right)}\left(W^{a}\mathbf{f}_{a}\right)-\mathbf{e}^{a}\mathbf{f}_{a}\label{eq:MixedDil}
\end{eqnarray}
\endnumparts

\numparts

\subsubsection*{Momentum space:}

\begin{eqnarray}
\frac{1}{2}\Omega_{\; b}^{a\;\; cd}\mathbf{f}_{c}\mathbf{f}_{d} & = & \mathbf{d}^{\left(y\right)}\boldsymbol{\gamma}_{\; b}^{a}-\boldsymbol{\gamma}_{\; b}^{c}\boldsymbol{\gamma}_{\; c}^{a}+\mathbf{D}\boldsymbol{\rho}_{\; b}^{a}-\boldsymbol{\rho}_{\; b}^{c}\boldsymbol{\rho}_{\; c}^{a}+k\Delta_{eb}^{ac}\eta^{ed}\mathbf{f}_{c}\mathbf{f}_{d}\label{eq:MomCurv}\\
\frac{1}{2}S_{a}^{\; bc}\mathbf{f}_{b}\mathbf{f}_{c} & = & \mathbf{d}^{\left(y\right)}\mathbf{f}_{a}-\boldsymbol{\gamma}_{\; a}^{b}\mathbf{f}_{b}-\frac{1}{2}\eta^{cb}\mathbf{d}^{\left(y\right)}\eta_{ac}\mathbf{f}_{b}\label{eq:MomBasis}\\
\frac{1}{2}T^{abc}\mathbf{f}_{b}\mathbf{f}_{c} & = & -k\eta^{ac}\left(\boldsymbol{\rho}_{\; c}^{b}\mathbf{f}_{b}-W^{b}\mathbf{f}_{b}\mathbf{f}_{c}-\frac{1}{2}\eta^{bd}\mathbf{d}^{\left(y\right)}\eta_{cd}\mathbf{f}_{b}\right)\label{eq:MomInv-1}\\
\frac{1}{2}\Omega^{bc}\mathbf{f}_{b}\mathbf{f}_{c} & = & \mathbf{d}^{\left(y\right)}\left(W^{a}\mathbf{f}_{a}\right)\label{eq:MomDil}
\end{eqnarray}
\endnumparts

\section*{Appendix 2: Coordinate to orthonormal basis}

The Euclidean and Lorentzian metric components are related in the
orthonormal basis by:

\begin{eqnarray*}
\eta_{ab} & = & s^{2}\left(\delta_{ab}-\frac{2}{s^{2}}s_{a}s_{b}\right)\\
\eta^{ab} & = & \frac{1}{s^{2}}\left(\delta^{ab}-\frac{2}{s^{2}}\delta^{ac}\delta^{bd}s_{c}s_{d}\right)\\
\delta_{ab} & = & \frac{1}{s^{2}}\left(\eta_{ab}+2s_{a}s_{b}\right)\\
\delta^{ab} & = & s^{2}\left(\eta^{ab}+2\eta^{ac}s_{c}\eta^{ad}s_{d}\right)
\end{eqnarray*}
where $s^{2}=\delta^{\alpha\beta}s_{\alpha}s_{\beta}>0$.

\section*{Appendix 3: Symmetric projection of the derivative of the solder
form}

For the calculation of the symmetric pieces of the connection, we
need to express the symmetric part, $\Xi_{cb}^{ad}h_{d}^{\;\alpha}\mathbf{d}h_{\alpha}^{\; c}$,
in terms of the metric. Expanding the metric derivatives,
\begin{eqnarray*}
k^{\alpha\mu}\mathbf{d}k_{\mu\beta} & = & k^{\alpha\mu}\mathbf{d}\left(h_{\mu}^{\; a}h_{\beta}^{\; b}\eta_{ab}\right)\\
 & = & h_{c}^{\;\alpha}h_{d}^{\;\mu}\eta^{cd}\left(\mathbf{d}h_{\mu}^{\; a}h_{\beta}^{\; b}\eta_{ab}+h_{\mu}^{\; a}\mathbf{d}h_{\beta}^{\; b}\eta_{ab}\right)\\
 & = & h_{c}^{\;\alpha}h_{d}^{\;\mu}h_{\beta}^{\; b}\eta^{cd}\eta_{ab}\mathbf{d}h_{\mu}^{\; a}+h_{b}^{\;\alpha}\mathbf{d}h_{\beta}^{\; b}\\
 & = & h_{c}^{\;\alpha}h_{\beta}^{\; b}\eta^{cd}\eta_{ab}\left(h_{d}^{\;\mu}\mathbf{d}h_{\mu}^{\; a}\right)+h_{b}^{\;\alpha}h_{\beta}^{\; c}\left(h_{c}^{\;\mu}\mathbf{d}h_{\mu}^{\; b}\right)\\
 & = & 2h_{c}^{\;\alpha}h_{\beta}^{\; b}\Xi_{ab}^{cd}\left(h_{d}^{\;\mu}\mathbf{d}h_{\mu}^{\; a}\right)
\end{eqnarray*}
so that we can write $\Xi_{cb}^{ad}\left(h_{d}^{\;\mu}\mathbf{d}h_{\mu}^{\; c}\right)$
explicitly, 
\begin{eqnarray*}
\Xi_{cb}^{ad}\left(h_{d}^{\;\mu}\mathbf{d}h_{\mu}^{\; c}\right) & = & \frac{1}{2}h_{\alpha}^{\; a}h_{b}^{\;\beta}k^{\alpha\mu}\mathbf{d}k_{\mu\beta}\\
 & = & \frac{1}{2}h_{\alpha}^{\; a}h_{b}^{\;\beta}k^{\alpha\mu}\mathbf{d}\left(s^{2}\delta_{\mu\beta}-2s_{\mu}s_{\beta}\right)\\
 & = & h_{\alpha}^{\; a}h_{b}^{\;\beta}\frac{1}{s^{2}}\left(\delta_{\beta}^{\alpha}\delta^{\nu\rho}s_{\rho}-\delta^{\alpha\nu}s_{\beta}+\delta_{\beta}^{\nu}\delta^{\alpha\mu}s_{\mu}\right)\mathbf{d}s_{\nu}\\
 & = & \frac{1}{s^{2}}\left(\delta_{b}^{a}\delta^{cd}s_{d}-\delta^{ac}s_{b}+\delta_{b}^{c}\delta^{ad}s_{d}\right)h_{c}^{\;\nu}\mathbf{d}s_{\nu}\\
 & = & -\frac{\left(1-k\beta^{2}\right)}{2\beta}\left(\delta_{b}^{a}\eta^{cd}s_{d}+\delta_{b}^{c}\eta^{ad}s_{d}+\eta^{ac}s_{b}+2\eta^{af}\eta^{ce}s_{b}s_{e}s_{f}\right)\eta_{cf}\mathbf{e}^{f}\\
 &  & -\frac{k\left(1+k\beta^{2}\right)}{2\beta}\left(\delta_{b}^{a}\eta^{cd}s_{d}+\delta_{b}^{c}\eta^{ad}s_{d}+\eta^{ac}s_{b}+2\eta^{af}\eta^{ce}s_{b}s_{e}s_{f}\right)\mathbf{f}_{c} 
\end{eqnarray*}

\bibliographystyle{unsrt}

\bibliography{/Users/jeffrey/Documents/PhD/Dissertation/MyDissertation/Dissertation}

\end{document}